\documentstyle[aps,preprint,psfig]{revtex}
\tightenlines
\draft
\widetext
\newcommand{\be}{\begin{equation}}
\newcommand{\bea}{\begin{eqnarray}}
\newcommand{\ee}{\end{equation}}
\newcommand{\eea}{\end{eqnarray}}
\renewcommand{\/}{\over}

\renewcommand{\Im}{{\rm Im}\,}
\newcommand{\Tr}{{\rm Tr}\,}

\begin{document}
\title{Universal spectral properties of spatially periodic quantum
systems with chaotic classical dynamics}

\author{T. Dittrich$^1$, B. Mehlig$^1$, H. Schanz$^1$ and U. Smilansky$^2$}
\address{
$\mbox{}^1$~Max-Planck-Institut f\"ur Physik komplexer Systeme, 
Bayreuther Str. 40, Haus 16\\ 
01\,187 Dresden, Germany\\
$\mbox{}^2$~Department of Physics of Complex Systems, 
The Weizmann Institute of Science\\ 
Rehovot 76\,100, Israel}

\date{\today}

\maketitle
\thispagestyle{empty}

\begin{abstract} 
We consider a quasi one-dimensional chain of $N$ chaotic scattering elements
with periodic boundary conditions.  The classical dynamics of this system is
dominated by diffusion.  The quantum theory, on the other hand, depends
crucially on whether the chain is disordered or invariant under lattice
translations.  In the disordered case, the spectrum is
dominated by Anderson localization whereas in the periodic case, the spectrum is
arranged in bands. We investigate the special features in the spectral
statistics for a periodic chain. For finite $N$, we define spectral form factors
involving correlations both for identical and non-identical Bloch numbers.
The short-time regime is treated
within the semiclassical approximation, where the spectral form factor can be
expressed in terms of a coarse-grained classical propagator which obeys a
diffusion equation with periodic boundary conditions. In the long-time regime,
the form factor decays algebraically towards an asymptotic constant. 
In the limit $N\rightarrow\infty$, we derive a universal scaling function
for the form factor.
The
theory is supported by numerical results for quasi one-dimensional periodic
chains of coupled Sinai billiards.
\end{abstract}


%%%%%%%%%%%%%%%%%%%%%%%%%%%%%%%%%%%%%%%%%%%%%%%%%%%%%%%%%%%%%%%%%%%%%%%%
%
% Introduction
%
%%%%%%%%%%%%%%%%%%%%%%%%%%%%%%%%%%%%%%%%%%%%%%%%%%%%%%%%%%%%%%%%%%%%%%%%
\section{Introduction}
\noindent
The goal of quantum chaology is to identify universal features which prevail
in systems whose classical dynamics is chaotic, and to relate the quantum
observables to their classical counterparts. Spectral statistics is one of the
topics where this approach has met with success, and the connection
between the spectral form factor (the Fourier transform of the two-point
correlation function) and the classical probability to stay, is a very
fruitful junction of mesoscopic physics and quantum
chaos~\cite{Ber85,HdA84,DiS95,AIS93,DSD92,AKL95}. In the present article
we develop the corresponding theory for extended, quasi one-dimensional
systems.

We consider composite systems consisting of $N$ chaotic unit cells of
length $a$ with periodic boundary conditions. The example we have in
mind is a chain of interconnected chaotic billiards, as shown in
Fig.~\ref{fig:chains}. In practice one can realize such systems by
constructing a mesoscopic ring with a circumference $L=Na$ which is
smaller than the dephasing mean free path, but much larger than the
elastic mean free path. The particle moves as a random walker between
the unit cells, which results on average in a diffusive evolution. It is
important to note from the outset that the diffusive nature of the
classical evolution is completely indifferent to the translational
symmetry of the ring --- whether it is periodic or disordered. In
contrast, quantum mechanics depends crucially on this feature:
disordered rings are affected by Anderson localization and in the limit
$N\to\infty$ the spectrum remains discrete with a Poissonian spectral
statistics \cite{DiS95}. Chains which are invariant under discrete
lattice translations develop continuous band spectra in the same limit.
Can one account for these two different quantum situations within a
theory that relies on the classical diffusive evolution which in itself
is indifferent to the symmetry? The present work provides an affirmative
answer to this question.  We study in particular the spectral form
factor for the periodic case, and show that in the limit $N \to
\infty$, it approaches a limit distribution which expresses the effect
of the levels clustering into bands.  

Let us consider the two scenarios
in some more detail:
In the {\em disordered} case, once the overall length $L=Na$ is sufficiently
larger than the Anderson localization length $\xi$, one can consider the
system qualitatively as a union of $L / \xi$ uncoupled systems. Within this
picture, the spectrum is a superposition of uncorrelated spectra, resulting in
a spectral form factor $K_N(\tau) \simeq K_{1}(\tau L /\xi)$, where the suffix
$1$ denotes the spectral form factor for the unit cell, as opposed to the form
factor $K_N(\tau)$ for the composite system.

In the {\em periodic} case, the symmetry with respect to discrete
lattice translations implies the existence of discretized energy bands.
According to Bloch's theorem, the energies are labeled by a
quasi-momentum (Bloch number) $q$ which takes discrete values for a
finite chain. 
Energy levels with different Bloch numbers are correlated over some
range in $q$.  Hence, on the time scale corresponding to an
energy scale of the order of the inter-band spacing, the spectrum looks
as if it were degenerate with a multiplicity $\sim N$. This corresponds
to a peak $\sim N$ in the form factor $K_N(\tau)$ at $\tau \sim 1/N$.  For
larger times, finer structures in the spectrum are resolved, until for
very large times ($\tau \stackrel{>}{\sim}1$), the discrete nature of
the spectrum is the dominant feature.  On this time scale $K_N(\tau)
\simeq 1$ (in the absence of symmetries). As will be shown in this
paper, the interpolation between the two time scales is given by a
power-law decay $ K_N(\tau ) \sim 1 /\tau $.  

Since we are dealing here with the periodic case, we make use of the extra
parameter which quantum mechanics provides, and consider spectral statistics
which test the correlations between spectra of different Bloch numbers. We
develop a semiclassical theory for these spectral measures and show that
they are intimately connected with the distribution of winding numbers in the
symmetry-reduced classical system, which in the present case is the unit cell
with periodic boundary conditions. This analysis makes use of the general
formalism proposed by Robbins \cite{Rob89}, and can be considered as an
extension of the previous work on band spectra in chaotic systems
\cite{LM94}.

It is important to emphasize that the considerations given above are rather
general and do not depend on the details of the actual system. To check the
universality of our arguments we have performed a parallel study where the system
of interest is not a billiard chain, but an appropriately defined version of
the quantum kicked rotor \cite{CCIF83}. This work will be reported elsewhere
\cite{DMSS96a}. The results obtained for the two models agree in all details,
giving credence to our claim that the spectral statistics for periodic
diffusive chains are universal.

The paper is organized as follows.  In section \ref{sec:qm} we derive a
secular equation for the energy spectrum of the billiard chain. It is based on
the scattering approach to quantization \cite{DS92b} and used here to compute
the spectrum numerically. The quantities allowing us to study spectral 
correlations for periodic quantum 
systems are introduced in section \ref{sec:corr}.  Section \ref{sec:scl} 
is devoted to the semiclassical theory of the form factor
for small times ($\tau\stackrel{<}{\sim}1/N$), while in section
\ref{sec:scaling} we discuss the form factor for 
intermediate times ($1/N \stackrel{<}{\sim} \tau < 1$). Finally,
in section \ref{sec:conclusions}, we summarize our results.

%%%%%%%%%%%%%%%%%%%%%%%%%%%%%%%%%%%%%%%%%%%%%%%%%%%%%%%%%%%%%%%%%%%%%%%%
%
% Quantum mechanics of periodic systems
%
%%%%%%%%%%%%%%%%%%%%%%%%%%%%%%%%%%%%%%%%%%%%%%%%%%%%%%%%%%%%%%%%%%%%%%%%
\section{Quantum mechanics}
\label{sec:qm}
\begin{figure}
\centerline{\psfig{figure=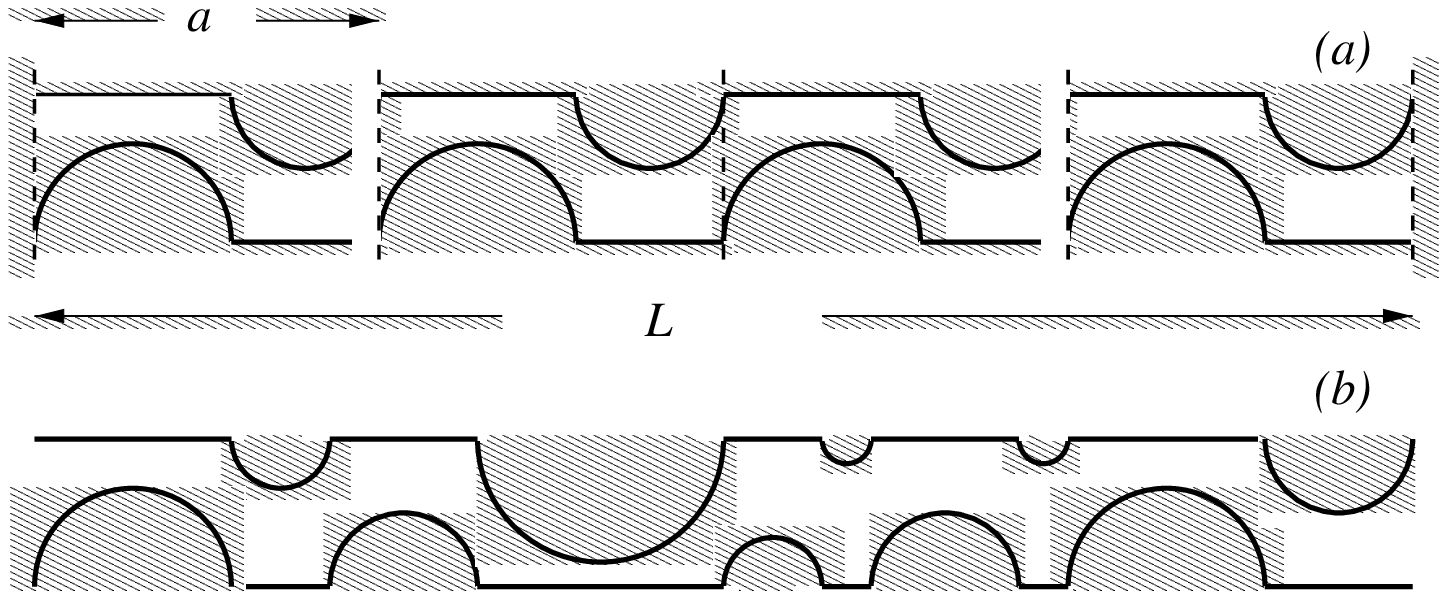,width=15cm}}
\caption{\label{fig:chains} Quasi one-dimensional chains of coupled
billiards. (a) shows a periodic chain while (b) shows a disordered chain. In
both cases periodic boundary conditions are imposed at the ends of the
chains.}
\end{figure}

\begin{figure}
\centerline{\psfig{figure=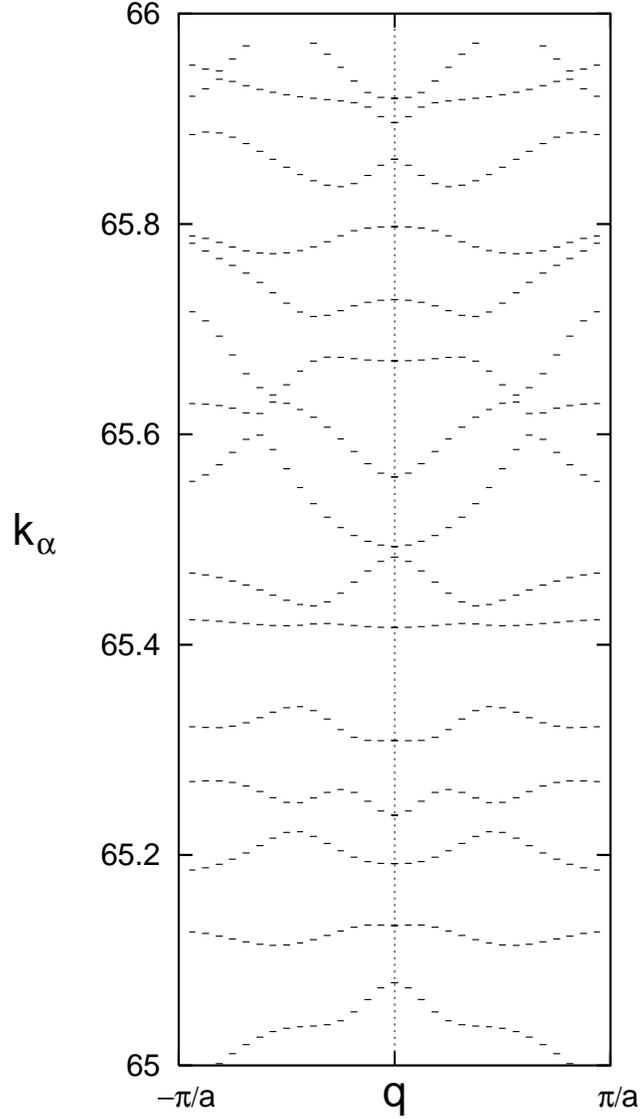,angle=270,height=15cm}}
\caption{\label{fig:spag}  Typical band spectrum of
a periodic billiard chain with $N=32$ unit cells. Shown are the levels
$k_{\alpha}(q)$ as a function of the Bloch number $q$ in the first Brillouin
zone, $-\pi/a \leq q < \pi/a$.  
}
\end{figure}

\noindent
In the present section we describe the quantization of the system shown in
Fig.~\ref{fig:chains}(a).  It is composed of $N$ identical unit cells of
length $a$ which form a quasi one-dimensional chain of total length $L=Na$.
Each unit cell represents a waveguide of unit width with two half-discs 
as shown in Fig.~\ref{fig:unitcell}. The radii are denoted by $R_-$
and $R_+$, respectively.
The wave function $\psi(x,y)$ satisfies the Schr\"odinger equation
\begin{equation}\label{helmholtz}
-\frac{\hbar^2}{2m}\Delta \psi(x, y)= E\psi(x, y)
\end{equation}
with Dirichlet boundary conditions on the walls of the waveguide and periodic
boundary conditions in $x$-direction
\be\label{pbc}
\psi(x+L, y)=\psi(x, y)\,.
\ee
\noindent  Moreover, as shown in Fig.~\ref{fig:chains}(a),
the system is invariant under discrete lattice translations $x\rightarrow
x+a$.  In conjunction with (\ref{pbc}), the translational symmetry
implies the existence of discretized energy bands $E_\alpha(q)$.
The energies are labeled by a discrete set of Bloch numbers
\begin{equation}
\label{qmom}
q={2\pi\nu\/L}\,,\qquad \nu=0,\dots,N-1
\end{equation}
\noindent in the first Brillouin zone, corresponding to the
$N$ irreducible representations of the group of lattice translations.
A typical band spectrum $E_\alpha(q)$ is shown in Fig.~\ref{fig:spag}.
The wave functions can be classified according to their Bloch numbers
and can be written as 
\be\label{bloch}
\psi_q(x,y)=\exp(iqx)\;u_q(x,y)\,,
\ee
\noindent where $u_q(x,y)$ is periodic in $x$ with period $a$ (Bloch's
theorem). 
Hence, for every Bloch number $q$ one has the quantization condition
\begin{eqnarray}
-\frac{\hbar^2}{2m}\Delta \,\psi_q(x,y)  &=&E(q) \,\psi_q(x,y)\,,\nonumber\\
\psi_q(x+a,y) &=& \exp(iqa)\;\psi_q(x,y)\;.
\label{eq:quantize}
\end{eqnarray}
\noindent Obviously, the spectrum $E_\alpha(q)$ has the following symmetries
\begin{eqnarray}
E_\alpha(q+2\pi/a) &=& E_\alpha(q)\,,\\
E_\alpha(-q) &=& E_\alpha(q)\,.
\label{eq:sym}
\end{eqnarray}
In what follows we discuss an implementation of the quantization condition
(\ref{eq:quantize}). Due to the translational invariance, it is sufficient to
quantize a single unit cell with periodic boundary conditions including an
additional phase $\exp(iqa)$, as shown in Fig.~\ref{fig:unitcell}. In the
straight-channel sections of the unit cell, the wave function can be
decomposed into normal modes \be\label{psi} \psi_q(x, y)=\sum_{j=1}^\infty
{\phi_{j}(y)\/\sqrt{k_j}} \left( a_{js}^{(+)} e^{+isk_j(x-x_s)} + a_{js}^{(-)}
e^{-isk_j(x-x_s)} \right)\,, \ee where \be \phi_j(y)=\sqrt{2}\;\sin(j\pi y),
\ee and the index $s=\mp$ designates the l.h.s.\ and r.h.s.\ of the unit cell,
respectively.  The index $n$ labels the normal modes, with wave numbers $k_n$
defined according to
\begin{equation}\label{kdef}
k_j=\left\{
\begin{array}{ll}
\;\sqrt{k^2-(j\pi)^2} & \qquad\mbox{for open modes\hspace{2mm} 
                                    ($j \le \Lambda(E)$)} \\ 
i\sqrt{(j\pi)^2-k^2}&\qquad\mbox{for closed modes 
                                    ($j > \Lambda(E)$)}\,,
\end{array}
\right.
\end{equation} 
where $\hbar k=\sqrt{2 m E}$.
The number of open modes $\Lambda(E)$ is given by the integer
part of $k/\pi$. For an open mode, $a_{js}^{(-)}$ represents the amplitude
of the incoming and $a_{js}^{(+)}$ that of the outgoing partial wave.  As the
semicircular obstacles are approached, the partial waves for $j>\Lambda(E)$
decrease (increase) exponentially. 
The coefficients of the
partial waves to the left and to the right of the obstacles
can be related to each other using Eq.~(\ref{eq:quantize}). One has
$a_{j,+}^{(\pm)} =e^{iqa}a_{j,-}^{(\mp)}$,
which can be written as a matrix equation
\be\label{qcond}
\pmatrix{a^{(-)}_-\cr a^{(-)}_+}=\Gamma(q)\,\pmatrix{a^{(+)}_-\cr a^{(+)}_+}
\ee
with
\be
\Gamma_{js,j^\prime s^\prime}(q)=\delta_{j,j^\prime}\,
                                 \delta_{s,-s^\prime}\,e^{isqa}\,.
\ee
We will now express the condition (\ref{qcond}) in terms of the generalized
scattering matrix $S(E)$ of the obstacle
\cite{DS92b} which provides a convenient starting point for both
the numerical implementation and the semiclassical theory. For this
purpose, we consider the Green function in the unit cell which
satisfies $[\Delta + k^2]\, G ({\bf r},{\bf r}^\prime) = -\delta ({\bf
r}-{\bf r}^\prime)$, with Dirichlet boundary conditions on the walls of the
waveguide and outgoing boundary conditions at the borders of the unit
cell. In the straight-channel sections, $G({\bf r},\bf{r}^\prime)$ 
can be written in the normal-mode
decomposition
\bea\label{gf}
G(x,y;x^\prime,y^\prime)={i\/2}
%\sum_{\stackrel{n,n^\prime}{s,s^\prime}}
\sum_{j,j^\prime}
{\phi_{j}(y)\/\sqrt{k_j}}{\phi_{j^\prime}(y')\/\sqrt{k_{j^\prime}}}
\left(\delta_{j,j^\prime}\,
      \delta_{s,s^\prime}\,e^{ik_j|x-x^\prime|}\right.\nonumber\\
\left.+S_{js,j^\prime s^\prime}
      \,e^{isk_j(x-x_s)+is'k_{j^\prime}(x^\prime-x_{s^\prime})}\right)\,.
\eea
The matrix $S$ depends on the particular geometry and remains to be
determined. Using (\ref{psi}), (\ref{gf}) and Green's theorem for the
unit cell it is straightforward to show that $S$ transforms the
amplitudes of the incoming partial waves into those of the outgoing
ones,
\be\label{strafo}
\pmatrix{a^{(+)}_-\cr a^{(+)}_+}=S\,\pmatrix{a^{(-)}_-\cr a^{(-)}_+}\,,
\ee
for any solution of (\ref{helmholtz}).  Hence, it can be regarded as a
generalized scattering matrix for the obstacle in the waveguide.
The usual unitary $S$-matrix is obtained by restricting $n$ and $n'$ to open
modes. We note that $S$ may be written in block form
\be\label{srt}
S=\pmatrix{r & t' \cr t & r'}\,,
\ee
in terms of the reflection and transmission matrices $r,r'$ and $t,t'$,
respectively. The phase shifts are defined as in Ref.~\cite{CMD88} such that
$S_{js,j^\prime s^\prime}=
\delta_{j,j^\prime}\,\delta_{s,-s^\prime}\,e^{ik_ja}$ for an empty unit cell
of length $a$.
Using (\ref{qcond}) to eliminate the coefficients of the outgoing 
waves in (\ref{strafo}) we find 
\be
[1-\Gamma(q)\,S(E)]\,\pmatrix{a^{(-)}_-\cr a^{(-)}_+}=0\,,
\ee
which can be satisfied by some combination of incoming waves provided
that
\be\label{eq:seceq}
\det[1-\Gamma(q)\,S(E)]=0\,.
\ee
This is the secular equation for the spectrum within the scattering approach
to quantization \cite{DS92b,SS95}. Together with the method described in
appendix~\ref{sec:smat} for the computation of the matrix $S$,
Eq.~(\ref{eq:seceq}) can be used to obtain the energy spectrum of the periodic
chain of scatterers numerically. For this purpose one restricts $S$ to a
finite matrix. 
In order to obtain well-converged eigenvalues,
at least four closed modes have to be taken into
account in the present case. 

%%%%%%%%%%%%%%%%%%%%%%%%%%%%%%%%%%%%%%%%%%%%%%%%%%%%%%%%%%%%%
%
% Spectral correlations in periodic systems
%
%%%%%%%%%%%%%%%%%%%%%%%%%%%%%%%%%%%%%%%%%%%%%%%%%%%%%%%%%%%%%
\section{Spectral correlations in periodic systems}
\label{sec:corr}
\begin{figure}
\centerline{\psfig{figure=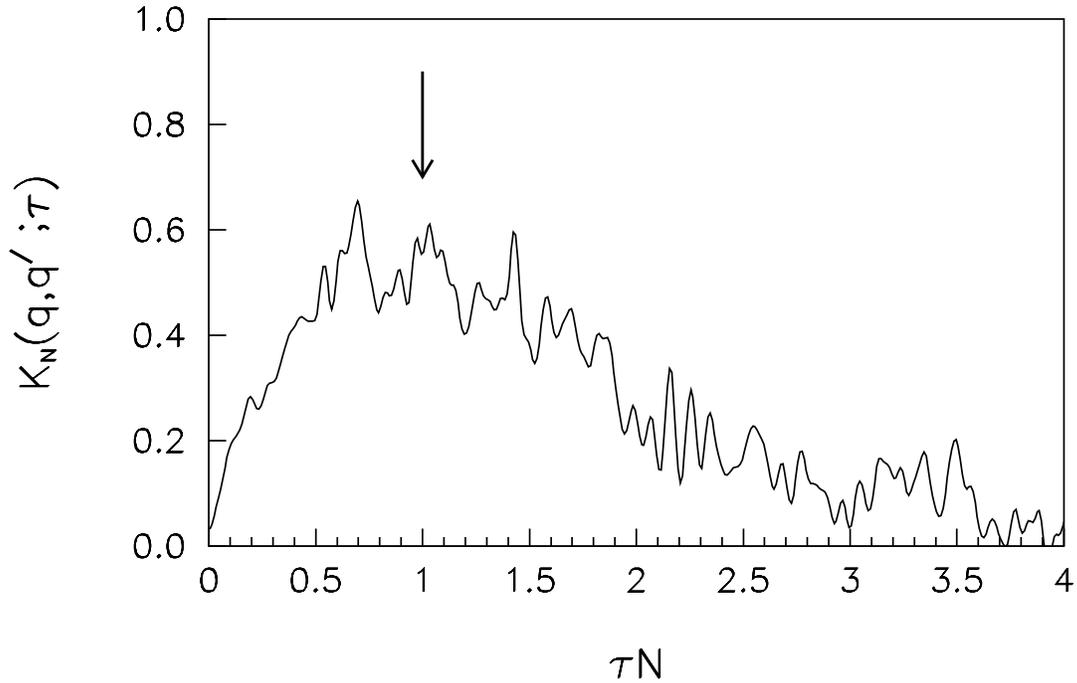,width=15cm}}
\caption{\label{fig:cross} Spectral form factor $K_N(q,q^\prime;\tau)$
for $q\neq q^\prime$ as a function of $\tau N$.
For large times $\tau>1/N$, $K_N(q,q^\prime;\tau)$ tends to zero since
levels of the same band decorrelate for $q\neq q^\prime$.
The arrow indicates $\tau\sim1/N$, a time of the
order of the inverse inter-band spacing.
The data are taken from a chain with $N=128$ unit cells.}
\end{figure}

\noindent
In the present section we discuss the spectral properties of
quasi one-dimensional periodic systems.  A typical spectrum is shown in
Fig.~\ref{fig:spag} (for $N=32$).  The spectral correlations are expressed in
terms of the density of states
\begin{equation}
\label{eq:dos}
d(E,q) = \sum_\alpha \delta(E-E_\alpha(q))\,.
\end{equation}
\noindent It is convenient to decompose the density
of states into a mean and an oscillatory part,
$d = \langle d\rangle + \widetilde{d}$. Note that $\langle d\rangle$
is the mean density of states for a given Bloch number. In other
words, it is the mean density of states
associated with one unit cell. 
In order to emphasize this, the mean density will henceforth be denoted
as $\langle d_1\rangle$. As usual, 
we unfold the spectrum~\cite{Boh89}
in order to obtain a unit total mean density. This is achieved by
\begin{equation}\label{eq:xdef}
x_\alpha(q) = N \langle d_1 \rangle E_\alpha(q)\,,
\end{equation}
\noindent where the energy dependence of the mean density of states
has been neglected. Note that the unfolding of the spectrum is done with
respect to the 
area of the entire chain~\cite{prl}
We define 
the unfolded density of states according to
\begin{equation}
d(x,q) = \sum_\alpha \delta(x-x_\alpha(q))\,.
\end{equation}
The spectral form factor $K_N(q,q^\prime;\tau)$ is then defined as 
\begin{equation}
K_N(q,q^\prime;\tau) = 
\Bigg\langle\int_{-\infty}^\infty \!{\rm d}\xi \;e^{-2\pi i\xi \tau}\,
\widetilde{d}(x+\xi/2,q)\,
\widetilde{d}(x-\xi/2,q^\prime)\Bigg\rangle\,.
\end{equation}
\noindent Here, $\tau$ is a rescaled time, related to  the physical time $t$ 
by $t = 2\pi\hbar \langle d_1\rangle N \tau$.  
For a finite spectral window the form factor can be
expressed as follows~\cite{Dit96}
\begin{eqnarray}
K_N(q,q^\prime;\tau) &=& \frac{N}{\Delta x}
\Bigg[
\sum_\alpha \;\;e^{2\pi i (x-x_\alpha(q))\tau} \;\,\;
        f(x\!-\!x_\alpha(q))\, -\widehat{f}(\tau)
\Bigg]\nonumber\\
&&\mbox{\hspace{0.15cm}}\times
\Bigg[
\sum_{\alpha^\prime} e^{-2\pi i (x-x_{\alpha^\prime}(q^\prime))\tau} 
f(x\!-\!x_{\alpha^\prime}(q^\prime)) -\widehat{f}(\tau)
\Bigg]\,.
\label{eq:Ktau}
\end{eqnarray}
Here, $f(x)$ is a spectral window function defined according to
\begin{equation}
\label{eq:23}
f(x) = 
\left \{
\begin{array}{ll}
1 & \mbox{for $|x| < \Delta x/2$}\,,\\
0 &\mbox{otherwise}\,.
\end{array}
\right .
\end{equation}
\noindent $\widehat{f}(\tau)$ denotes the Fourier transform of $f(x)$.
Note that $f(x)$ 
is normalized such that its integral equals the number $\Delta x$ of states 
contained in the spectral window,
\begin{displaymath}
\int_{-\infty}^\infty \!\!{\rm d}x \;f(x) = \Delta x\,.
\end{displaymath}
\noindent The form factor $K_N(q,q^\prime;\tau)$ is normalized such
that $K_N(q,q;\tau)\rightarrow 1$ for large $\tau$. In fact, for large $\tau$,
only the diagonal contributions to the double sum in Eq.~(\ref{eq:Ktau}) have
to be retained and one has (for $q$ and $q^\prime$ not related by the
band symmetries discussed in section~\ref{sec:qm})
\begin{eqnarray}
K_N(q,q^\prime;\tau) &\simeq& 
\delta_{q,q^\prime}
              \frac{N}{\Delta x} \sum_\alpha f^2(x-x_\alpha(q))\nonumber\\
&\simeq& \delta_{q,q^\prime}
              \frac{1}{\Delta x} \int_{-\infty}^\infty 
                     \!\!{\rm d}x\; f^2(x) \nonumber\\
&=& \delta_{q,q^\prime}\,. 
\label{eq:largetau}
\end{eqnarray}
\noindent The fact that the form factor tends to
a constant for $q=q^\prime$ and large $\tau$ is a
consequence of the discrete nature of the spectrum.
The corresponding quantum dynamics
is governed by quasi-periodic motion on
the time scale of the inverse intra-band spacing.
According to Eq.~(\ref{eq:largetau}),
correlations between different Bloch numbers 
decrease rapidly for large $\tau$,
provided the Bloch numbers are not related by symmetries.
An example is given in Fig.~\ref{fig:cross},
where $K_N(q,q^\prime;\tau)$ is shown for $q\neq q^\prime$. 
As expected, the form factor tends to zero for large $\tau$. 
On the other hand we observe significant cross correlations between
different Bloch numbers at intermediate values of $\tau$, of the
order of $\tau\simeq 1/N$. This time
scale corresponds to the mean inter-band spacing and the correlations are
associated with the existence of energy bands.  Before discussing these cross
correlations in detail, we consider the spectral correlations at fixed Bloch
number $q$.

\subsection{Spectral correlations at fixed Bloch number $q$}
\label{sec:qq}
\noindent
In the present subsection we discuss spectral correlations
for fixed Bloch number $q$. Consider for instance the
spectrum shown in Fig.~\ref{fig:spag}.
Since the classical dynamics of the unit cell is chaotic, the spectral
correlations at fixed $q$ are expected to be described by random-matrix
theory. The quantization condition (\ref{eq:quantize})
of section~\ref{sec:qm} can be rewritten as follows
\begin{eqnarray*}
-\frac{\hbar^2}{2m} \Big[\Big(\frac{\partial}{\partial x}+iq\Big)^2+
\frac{\partial^2}{\partial y^2}\Big]u_q(x,y) &\equiv& \widehat{H}(q) u_q(x,y)
= E(q) u_q(x,y)\nonumber\\
u_q(x+a,y) &=& u_q(x,y).
\end{eqnarray*}
\noindent Obviously, the Bloch number $q$ can be regarded as reflecting
an Aharonov-Bohm flux $\phi$, with
$q = 2\pi/a\;\phi/\phi_0$ and $\phi_0 = 2\pi\hbar/e$.
For $q=0$, the Hamiltonian $\widehat{H}(q)$ is invariant with respect to
time reversal and the spectral properties
are described by the ensemble of Gaussian orthogonal
random matrices (GOE). In the  case $q = \pi/a$,
$\widehat{H}(q)$ is invariant under the anti-unitary
transformation $\widehat{\theta} = \exp(-2iqx) \widehat{J}$, where
$\widehat{J}$ is the operator of complex conjugation. For $q=\pi/a$, 
the appropriate ensemble is hence also the
orthogonal one (GOE).  For $q=\pi/2a$, on the other hand, the spectral
properties are described by the ensemble of Gaussian unitary random matrices (GUE).
As $q$ changes from zero to $\pi/2a$, one expects a
crossover from orthogonal to unitary statistics.

\begin{figure}
\centerline{\psfig{figure=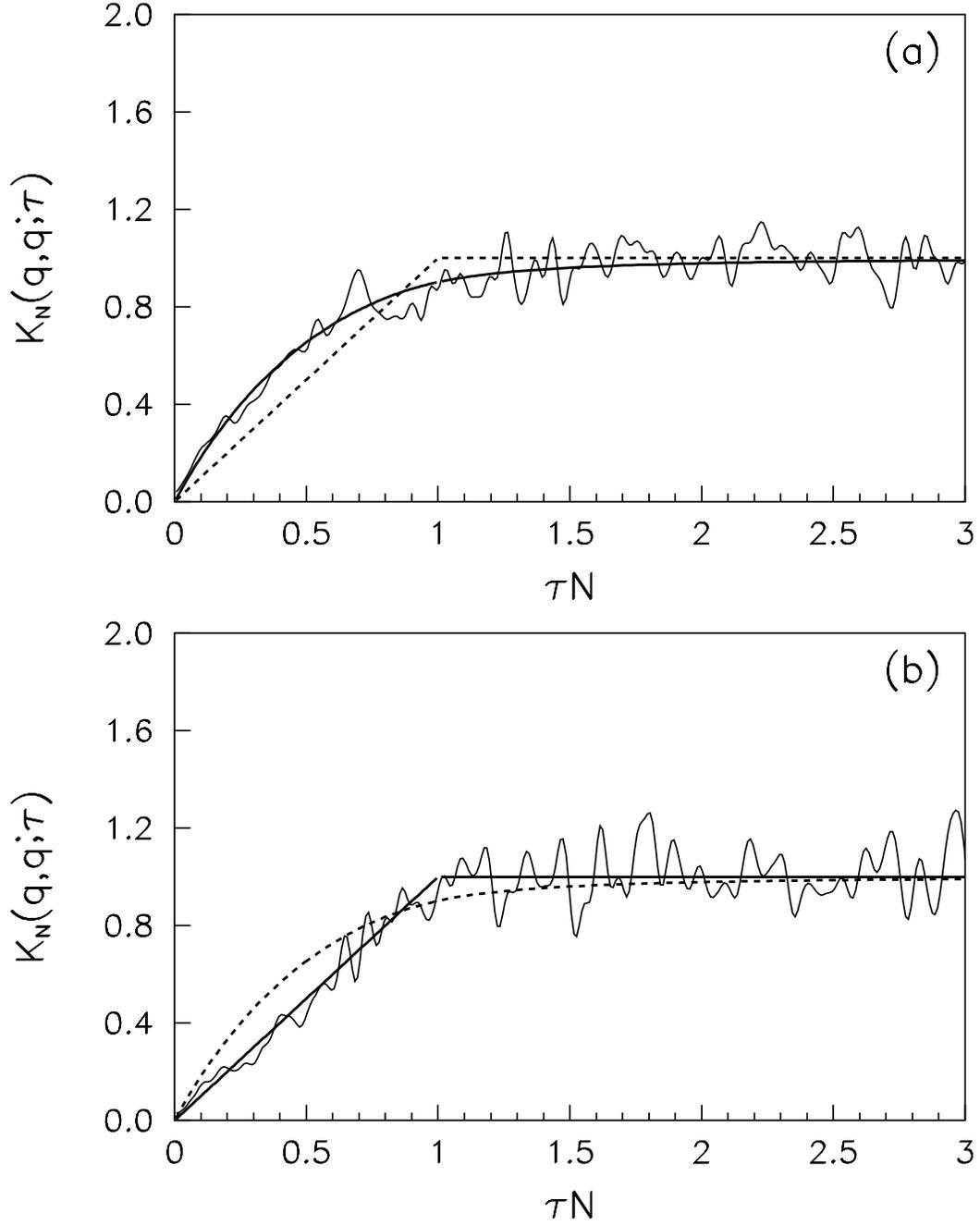,width=15cm}}
\caption{\label{fig:kt} Form factors $K_N(q,q;\tau)$ 
for different values of the Bloch number $q$ as
a function of $\tau N$. In (a) we show the case
$q=0$ where the spectrum shows GOE statistics, and in (b) the case $q=\pi/2a$
where the spectrum shows GUE statistics.}
\end{figure}

\begin{figure}
\centerline{\psfig{figure=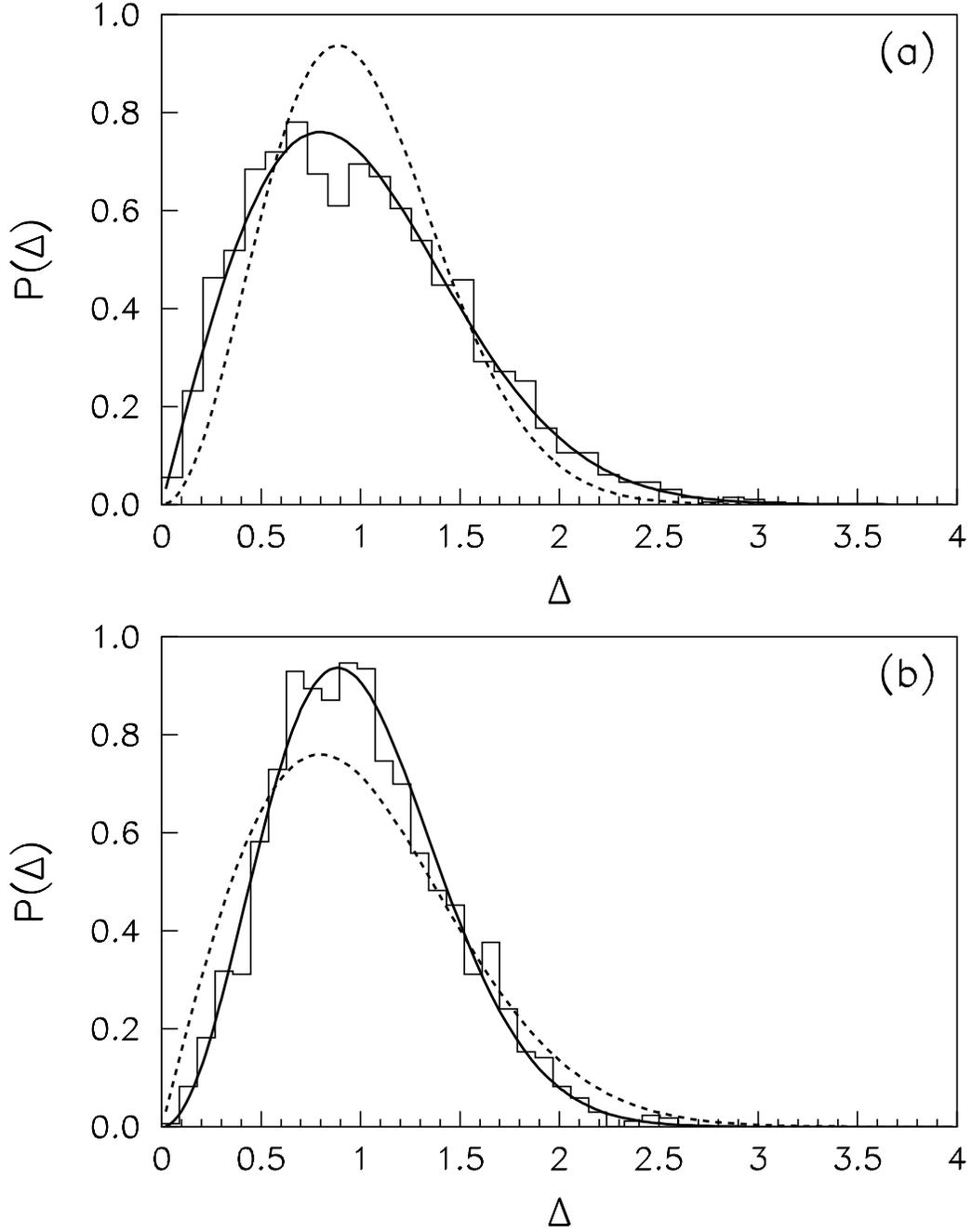,width=15cm}}
\caption{\label{fig:nn} Histograms $P(\Delta)$
of the nearest-neighbour spacings $\Delta$,
for fixed Bloch number $q$. (a) shows the
case $q=0$ where the spectrum shows GOE statistics, while
(b) shows the case $q=\pi/2a$, where the spectrum shows GUE statistics.}
\end{figure}

Fig.~\ref{fig:kt} shows the spectral form factor $K_N(q,q;\tau)$ as a function
of $\tau$ for two different values, $q=0$ and $q=\pi/2a$, of the Bloch number.
The universal results for the orthogonal and unitary ensembles \cite{Boh89}
are displayed with solid and dashed lines, respectively.  
As expected, we observe GOE
behaviour for $q=0$ and GUE behaviour for $q=\pi/2a$. Similarly, the histogram
$P(\Delta)$ of nearest-neighbour spacings 
exhibits a crossover from GOE to GUE statistics.
This is shown in Fig.~\ref{fig:nn}.
This crossover occurs on an $\hbar$-dependent scale. 
In the semiclassical limit $\hbar \rightarrow 0$,
the transition is instantaneous.

\subsection{Spectral cross correlations}
\label{sec:cross}
\noindent
In the present subsection we discuss correlations between
different Bloch numbers. These correlations
decay rapidly for large
$\tau$, as shown for instance in Fig.~\ref{fig:cross}.
However, it will turn out in the following that spectral cross correlations at
intermediate values of $\tau$ are of particular importance in extended
periodic quantum systems, since they are directly related 
to the existence of discretized energy bands. 
In order to analyse correlations between different values of $q$, we define
\begin{equation}
\label{eq:S1}
S_N(n;\tau) = 
\frac{1}{N} \sum_{q,q^\prime} e^{in(q-q^\prime)a} K_N(q,q^\prime;\tau)\,.
\end{equation}
\noindent Note that $S_N(0;\tau)$ is just the conventional spectral form factor
$K_N(\tau)$ for the composite system referred to in the introduction.
Rewriting $S_N(n;\tau)$ in terms of a Bloch-number specific density of states
\begin{equation}
\label{eq:dxn}
d(x,n) = \sum_q e^{inqa} \,d(x,q)\,,
\end{equation}
\noindent we obtain
\begin{equation}
\label{eq:S2}
S_N(n;\tau) = 
\Bigg\langle\int_{-\infty}^\infty \!{\rm d}\xi \;e^{-2\pi i\xi \tau}\;
\widetilde{d}(x+\xi/2,n)
\widetilde{d}(x-\xi/2,n)\Bigg\rangle\,.
\end{equation}
\noindent In the following, we discuss the long-time asymptote of $S_N(n;\tau)$.
For a finite spectral window, $S_N(n;\tau)$ can be expressed
in a form similar to Eq.~(\ref{eq:Ktau}) 
and for $\tau \gg 1/\Delta x$ one has
\begin{equation}
\label{eq:S3}
S_N(n;\tau) \simeq \frac{1}{\Delta x}
    \Bigg| \sum_q\sum_\alpha e^{inqa-2\pi i x_\alpha(q)\tau}\, 
    f(x\!-\!x_\alpha(q))\Bigg|^2\,.
\end{equation}
Taking into account the symmetry of the spectrum with respect
to $q\rightarrow 2\pi/a-q$ and in the absence of further symmetries, the
phases of the off-diagonal terms 
in the double sum are random and cancel. Using $q=2\pi\nu/L$, one has
\begin{eqnarray}
\phantom{S_N(n;\tau)}
&\simeq&  \frac{1}{N}\Big[2+4
    \sum_{\nu = 1}^{N/2-1}
    \cos^2\Big(\frac{2\pi\nu n}{N}\Big) \Big]
      \;\frac{1}{\Delta x}\int_{-\infty}^\infty \!{\rm d}x \;f^2(x),
\end{eqnarray}
\noindent assuming $N$ to be an even integer.
% By the normalization of $f(x)$ we obtain
Using $\int\!dx\;f^2(x) = \Delta x$ (compare Eq.~(\ref{eq:23})), we obtain
\begin{eqnarray}
\phantom{S_N(n;\tau)}
&=&
\left \{
\begin{array}{ll}
\displaystyle
2\big(1-\frac{1}{N}\big)
&\qquad\qquad\mbox{for $n=0,N/2$\,,}\\
\displaystyle
1-\frac{2}{N}            &\qquad\qquad\mbox{otherwise .}
\end{array}
\right .
\label{eq:asympt}
\end{eqnarray}
\noindent For large $N$, 
$S_N(n;\tau)\rightarrow 2$ for $n=0,N/2$.
This is a consequence of the band symmetry 
(\ref{eq:sym})
discussed in 
section~\ref{sec:qm}.

%%%%%%%%%%%%%%%%%%%%%%%%%%%%%%%%%%%%%%%%%%%%%%%%%%%
%
% Semiclassical analysis
%
%%%%%%%%%%%%%%%%%%%%%%%%%%%%%%%%%%%%%%%%%%%%%%%%%%
\section{Semiclassical analysis}
\label{sec:scl}
\noindent
In the following we obtain a semiclassical approximation for
$S_N(n;\tau)$. This is done in two steps. First, we derive a semiclassical
approximation for the density of states for a periodic quantum system. It will
be seen below that in the present case it is particularly convenient to start from
Eq.~(\ref{eq:seceq}). As discussed in Ref.~\cite{Rob89}, the resulting
semiclassical expression features the characters $\chi(n,q)=\exp(inqa)$ 
of the group of
lattice translations.  Second, we introduce the diagonal approximation for the
form factor.  The diagonal approximation 
is adequate for times smaller than
the Heisenberg time of the unit cell, that is for 
$t\stackrel{<}{\sim} t_H = 2\pi\hbar \langle d_1\rangle$ or
equivalently for $\tau \stackrel{<}{\sim} 1/N$. A theory covering
the long-time behaviour is
worked out in section~\ref{sec:scaling}.

\subsection{Semiclassical approximation for the density of states} 
\noindent
We now proceed to derive a semiclassical approximation for the energy
spectrum. The starting point is Eq.~(\ref{eq:seceq}), in which $S(E)$ is
restricted to the unitary $\Lambda\times\Lambda$-matrix corresponding to the
open modes. In this case the spectral density following from (\ref{eq:seceq})
may be written as \cite{DS92b}
\be\label{dos}
d(E,q)=\sum_{\alpha=1}^\infty \delta(E-E_{\alpha}(q))=
{{\rm d}\/{\rm d} E}{\Theta(E)\/2\pi}+
{{\rm d}\/{\rm d} E}\Im\sum_{m=1}^\infty \frac{1}{m\pi}
\Tr[\Gamma(q) S(E)]^m\,,
\ee
where $\Theta(E)=-i\log\det S(E)$ denotes the total phase of the
$S$-matrix. Using a theorem by M.\,G.\,Krein \cite{BY93}, it is possible
to obtain a semiclassical approximation for $\Theta$ which contains a
smooth part given by the generalized Weyl law \cite{BH76} for the
unit cell and an oscillating part which represents the standard
Gutzwiller contributions from all the periodic orbits which are
trapped inside the unit cell \cite{PSSU96b}.
In order to obtain a semiclassical approximation for $\Tr(\Gamma S)^m$,
we first use (\ref{gf}) at $x=x_s$, $x'=x_{s'}$ and find
\be
(\Gamma S)_{js,j^\prime s^\prime}=2s\,e^{isqa}\sqrt{k_{j^\prime}\/k_j}
\int_0^1\,{\rm d}y\,{\rm d}y'\,\phi_j(y)\phi_{j^\prime}(y')
\left.{\partial\/\partial x}G(xy;x_{s'}y')\right|_{x=x_{-s}}\,.
\ee
The completeness of the normal modes
$\sum_{j=1}^\infty\phi_j(y)\phi_j(y^\prime)=\delta(y-y^\prime)$ 
for $0<y,y^\prime<1$ allows
to express the powers of $\Gamma S$ in coordinate space as multiple
integrals over the Green function
\bea\label{trsn}
\Tr (\Gamma S)^m
=2^m e^{inqa}\sum_{s_1,\dots,s_m}\int\limits_0^{1} {\rm d}y_1\dots {\rm d}y_m\;
&&\left[s_1{\partial\/\partial x_1}\,G(x_1,y_1;x_{s_2},y_2)\right]_{x_1=x_{-s_1
}}
\nonumber \\
&\times&\cdots \nonumber\\ &\times&
\left[s_m{\partial\/\partial x_m}\,G(x_m,y_m;x_{s_1},y_1)\right]_{x_m=x_{-s_m}},
\eea
where the winding number $n$ is defined as
\be\label{wnum}
-n=s_1+\cdots+s_m\,.
\ee
The standard semiclassical expression for the Green function 
\be\label{gscl}
G(x,y;x',y')={1\/2}\sum_{\alpha}
\sqrt{i\/2\pi k}\left|\partial x_\bot\/\partial \varphi'\right|_{x'_\bot
}^{-1/2}
\exp\left({ikL_{\alpha}-i\,{\pi\/2}\mu_{\alpha}}\right)
\ee
can now be inserted into (\ref{trsn}). The sum in (\ref{gscl}) runs
over all classical trajectories $\alpha$  leading from $(x',y')$ to $(x,y)$.
$L_{\alpha}$ is the length of the path $\alpha$ and $\mu_{\alpha}$ contains the
Maslov index of the trajectory plus twice the number of reflections from
the walls of the waveguide. Due to the special geometry the Maslov
index happens to be zero for all trajectories.  $\varphi$ and $x_\bot$
denote the direction of the trajectory and the coordinate normal to
it, respectively.

\begin{figure}
\centerline{\psfig{figure=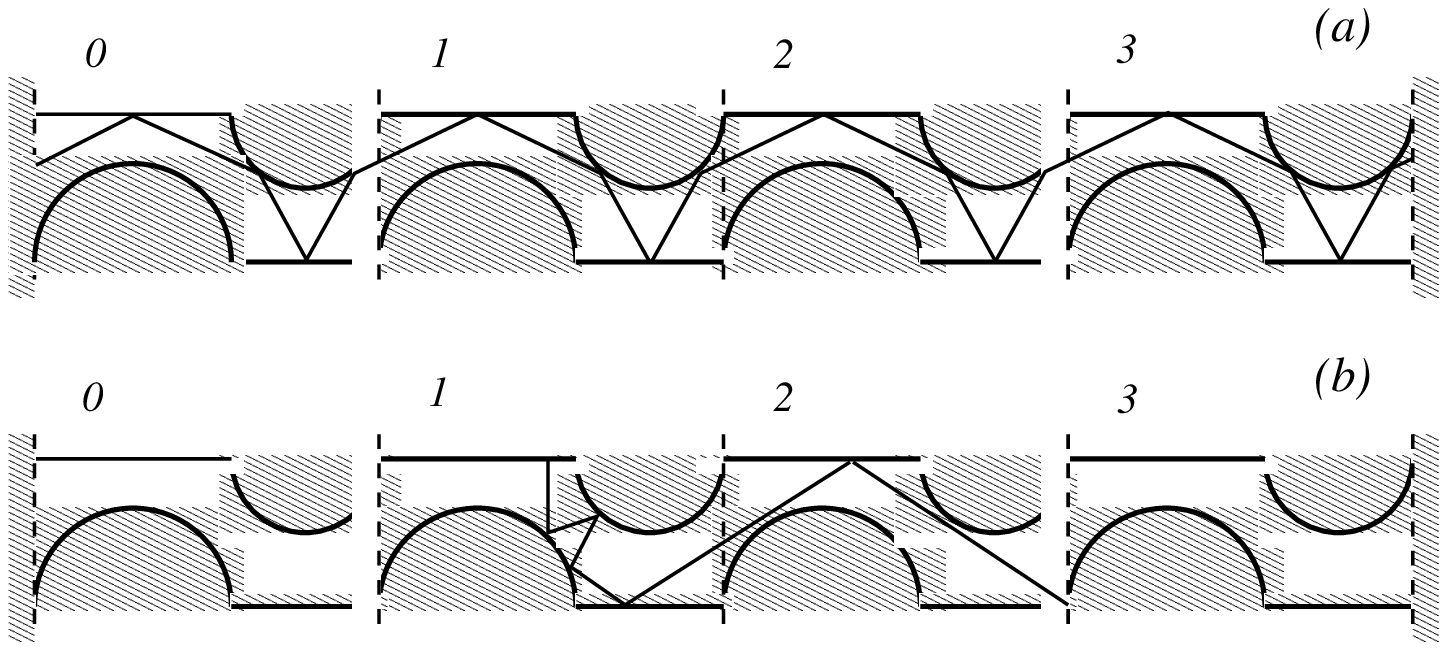,width=15cm}}
\caption{\label{fig:ext} Topology of periodic orbits in the
periodic chain. (a) shows a periodic orbit with
winding number $n=1$ while (b) shows a periodic orbit
with winding number $n=0$. Note that the periodic orbit
in (b) is self-retracing.}
\end{figure}

The integrals in (\ref{trsn}) are evaluated in the saddle-point approximation. 
The condition of stationary phase is satisfied for all
periodic orbits $p$ of the unit cell with periodic boundary conditions,
{\em i.e.}, whenever a periodic orbit leaves the unit cell
to the left or to the right, it is reinjected from
the opposite side.
$\Tr (\Gamma S)^m$ contains all those orbits for which
this happens exactly $n_p=m$ times, irrespective of which of
the two boundaries is hit. In contrast, $n_p$ defined in
(\ref{wnum}) counts the number of reinjections including a sign
according to the direction, such that $n_p$ finally
represents the number of windings of the periodic orbit around the
unit cell (see Fig.~\ref{fig:ext}).
Obviously the orbits contributing to the total phase have a
winding number $n_p=0$ and for the other orbits we have
$|n_p|<m_p$. We obtain 
\be\label{eq:dosq}
\widetilde{d}(E,q)=\frac{1}{2\pi\hbar}
\sum_{p,r} w_{p,r} T_p\;
%\exp\left(ikrL_p+irn_pqa-i{\pi\/2}r\mu_p\right)\,,
\exp\left(\frac{i}{\hbar} r S_p+irn_pqa-i{\pi\/2}r\mu_p\right)\,,
\ee
where $r = -\infty,\ldots,\infty$ (excluding $r=0$),
$p$ is summed over the
primitive periodic orbits,
$w_{p,r} = |\det(M_p^r-1)|^{-1/2}$,
$M={\partial(x_\bot,\varphi)/\partial(x'_\bot,\varphi')}$ is the stability
matrix of the periodic orbit $p$, and $S_p = \hbar k L_p$ is
the corresponding action.
The discrete translation symmetry results in an
additional phase which depends on the winding number of the
orbit. Eq.~(\ref{eq:dosq}) represents a special case of the general result
obtained in \cite{Rob89}.
According to Ref.~\cite{Rob89}, the additional phase factors are related
to the characters $\chi(n,q)$ of the irreducible representations
of the group of lattice translation. 

%%%%%%%%%%%%%%%%%%%%%%%%%%%%%%%%%%%%%%%%%%%%%%%%%%%%%%%%%%%%%%%%%%%%%%
\subsection{Diagonal approximation for $S_N(n,\tau)$}
\noindent
In the following we derive the semiclassical approximation for $S_N(n;\tau)$ in
the diagonal approximation~\cite{Ber91}.  
% Using
% Eq.~(\ref{eq:xdef}), we rewrite the semiclassical
% expression for the density of states (\ref{eq:dosq}) as
% \begin{displaymath}
% \widetilde{d}(x,q) = \frac{1}{2\pi\hbar N\langle d_1\rangle} 
% \sum_{p,r} \chi(rn_p,q)\, w_{p,r}\, T_p\, \exp\left(\frac{i}{\hbar} r S_p(x)
% +i\frac{\pi}{2} r\mu_p\right)\,.
% \end{displaymath}
% \noindent From (\ref{eq:dxn}) we find
Using Eqs.~(\ref{eq:xdef}) and (\ref{eq:dxn}), we have from (\ref{eq:dosq})
\begin{displaymath}
\widetilde{d}(x,n) = \frac{1}{2\pi\hbar\langle d_1\rangle}\sum_{p,r}
\delta_{n,rn_p}^{(N)} w_{p,r} T_p \exp\left(\frac{i}{\hbar} rS_p
+i\frac{\pi}{2} r\mu_p\right)\,.
\end{displaymath}
\noindent Here $\delta_{n,m}^{(N)}$ denotes a Kronecker delta
with the arguments taken modulo $N$. Expanding the
actions according to $S_p(E\pm \epsilon/2) \simeq S_p(E) \pm T_p\,\epsilon/2$,
we have
\begin{eqnarray*}
S_N(n;\tau) &=& 
\Bigg\langle\int_{-\infty}^\infty \!{\rm d}\xi \;e^{-2\pi i\xi \tau}\;
\widetilde{d}(x+\xi/2,n)
\widetilde{d}(x-\xi/2,n)\Bigg\rangle\\
&=& \frac{1}{(2\pi\hbar\langle d_1\rangle)^2}
\int_{-\infty}^\infty {\rm d}\xi e^{-2\pi i \xi\tau} 
\Bigg \langle
\sum_{p,r}\sum_{p^\prime,r^\prime} 
                 \delta_{n,r n_p}^{(N)} \delta_{n,r^\prime n_{p^\prime}}^{(N)}
                  w_{p,r} w_{p^\prime,r^\prime}^{\displaystyle\ast} 
                  \,T_p \,T_{p^\prime}\\
&&\hspace{2cm}
                 \times\exp\left(\frac{i}{\hbar}(r S_p - r^\prime S_{p^\prime})
                      + \frac{i}{2\hbar}(r T_p + r^\prime T_{p^\prime})\epsilon
                      + i\frac{\pi}{2}(r \mu_p - r^\prime \mu_{p^\prime})
\right)\Bigg\rangle\,,
\end{eqnarray*} 
\noindent where $\xi = N\langle d_1 \rangle \epsilon $.
Provided the actions $S_p$ are uncorrelated, 
the off-diagonal terms in the double sum over primitive
periodic orbits vanish
upon averaging.
This is certainly the case for
periodic orbits with periods less than some fraction of the Heisenberg time
of the unit cell, $T_p \stackrel{<}{\sim} t_H = 2\pi\hbar\langle d_1 \rangle$, which
restricts the applicability of the present approximation to times 
smaller than
the Heisenberg time of the unit cell, or equivalently, to 
$\tau \stackrel{<}{\sim} 1/N$.  
Furthermore we note that
pairs of
time-reversed orbits have in general opposite winding numbers. However, 
periodic orbits with $n_p=0$ or $N/2$ 
give rise to an additional degeneracy, since in these cases
$-n_p = n_p \;(\mbox{mod} N)$.
We define the coherence factor
\begin{equation}
\label{eq:g}
\gamma_n = 
\left\{
\begin{array}{ll}
2 & \qquad\mbox{for $n= 0,N/2$}\,,\\
1 & \qquad\mbox{otherwise\,,}
\end{array}
\right .
\end{equation}
\noindent 
and obtain 
\begin{eqnarray}
S_N(n;\tau) 
&\simeq& \gamma_n\tau N^2 \sum_{p,r}\,
                 \delta^{(N)}_{n,rn_p} \,
                 |w_{p,r}|^2 \,T_p \;
                 \delta\left(rT_p-2\pi\hbar\langle d_1\rangle N\tau\right)\,.
\label{eq:sump}
\end{eqnarray}
\noindent For $n=0$, the form
factor $S_N(0;\tau)$ is just the conventional spectral form factor
$K_N(\tau)$ for the composite system, {\em i.e.}, 
for a billiard chain with periodic boundary conditions~(\ref{pbc}). 
The primitive periodic orbits of the entire chain have winding numbers $n_p$ which
are multiples of $N$, as opposed to the periodic orbits of the unit cell.
The Kronecker delta in Eq.~(\ref{eq:sump}) ensures that for $n=0$, only
periodic orbits of the entire chain contribute. 

\begin{figure}
\centerline{\psfig{figure=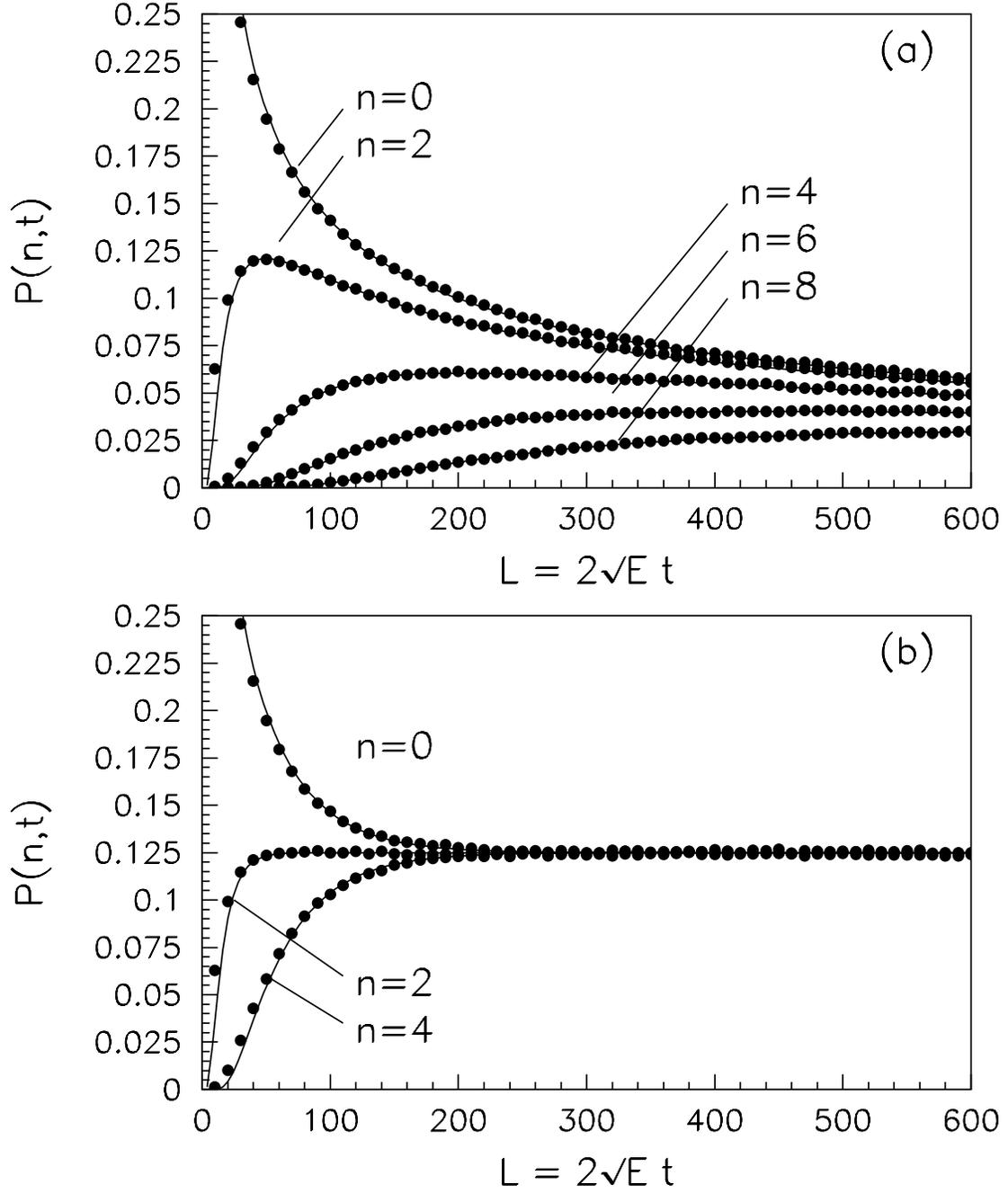,width=15cm}}
\caption{\label{fig:Pnt128} Diffusion propagator $P(n;t)$
for a periodic chain.
(a) shows the distribution of $P(n;t)$ for $N=128$,
(b) shows $P(n;t)$ for $N=8$, as a function of $l=\hbar kt/m$
(the energy dependence of the classical mechanics has been scaled out).
Note that for large $t$, $P(n;t)$ approaches $1/N$
(compare Eq.~(\protect\ref{eq:saturation})).} 
\end{figure}

The sum over periodic orbits in Eq.~(\ref{eq:sump}) can be evaluated
as follows.
According to appendix~C, it can be replaced by the
coarse-grained classical propagator $P(n;t)$ defined in Eqs.~(\ref{pfo})
and (\ref{pptr}).
Furthermore,
the classical dynamics in the periodic chain is
diffusive in the $x$-direction (provided direct trajectories
are eliminated). This is due to the chaotic scattering
in the unit cells. Under these circumstances, 
$P(n;t)$ is given by the expression (\ref{eq:Pn})
derived in appendix~C. In order to show that the classical dynamics in 
the billiard chain is adequatly described by Eq.~(\ref{eq:Pn}),
we have performed numerical simulations of the classical dynamics
in periodic chains. The chains consist of $N=8$ or $N=128$ copies
of the unit cell shown in Fig.~\ref{fig:unitcell}. The radii of 
the half discs have been chosen large enough so as to eliminate
direct bouncing-ball trajectories ($R_- = 0.8$ and
$R_+ = 0.4$ in units of the channel width). The lengths of the straight-channel
sections are $\sim 0.1$ in the same units. The results of the numerical
simulations are shown in Fig.~\ref{fig:Pnt128} (solid dots), 
as a function of the trajectory length $l = \hbar k t/m$.
For a given trajectory length $l$, we have
started $10^6$ trajectories and calculated the average probability to
move $n$ unit cells away from the starting point.
For convenience, the results are given as a function of lengths
$l$ rather than times $t$.
Also shown is a fit to 
Eq.~(\ref{eq:Pn}). The fitting parameter is the diffusion constant $D$.  
In the case shown, we obtain for the scaled diffusion constant
$D\, m/\hbar k a^2 \simeq 0.082$.  We observe good agreement between
the numerical simulations and the results of Eq.~(\ref{eq:Pn}).

Replacing the sum over periodic orbits in Eq.~(\ref{eq:sump}) by
the appropiate expression for the coarse-grained classical propagator,
we obtain (neglecting repetitions of periodic orbits,
since primitive periodic orbits proliferate exponentially)
\begin{eqnarray}
\label{eq:result}
S_N(n;\tau) &\simeq& \gamma_n \tau N^2 P(n;2\pi\hbar N\langle d_1\rangle\tau)
\nonumber \\
&\simeq& \gamma_n N \sqrt{N\tau\over 2g_1}\sum_\mu
\exp\left(-{\pi\over 2g_1\tau}{(n-\mu N)^2\over N}\right)\,,
\end{eqnarray}
where we have introduced the dimensionless parameter 
$$
g_1 = {2\pi^2\hbar\langle d_1\rangle D\over a^2}\,.
$$
The parameter $g_1$ is analogous to the dimensionless conductance
of disordered conductors~\cite{Dit96}. 
Note that it is defined with respect
to the unit cell.
However, since we do not assume static disorder within the unit cell,
the analogy is purely formal.

Eq.~(\ref{eq:result}) is the central result of this section.  It relates the
quantum spectral form factor $S_N(n;\tau)$ to the classical diffusion propagator
$P(n;t)$ of winding numbers $n$. 
According to Eq.~(\ref{eq:result}), the behaviour of $S_N(n;\tau)$ depends
crucially on the values of $N$ and $g_1$.
We distinguish two cases, 
$g_1/N^2 > 1$ and $g_1/N^2 < 1$. The spectral 
properties are qualitatively different in these two regimes.
For $g_1/N^2 > 1$, the classical dynamics becomes ergodic
before the energy-time uncertainty relation allows to resolve
the inter-band spacing. This corresponds to a time scale of the
order of the Heisenberg time of the unit cell, 
$t_{\rm H} = 2\pi\hbar \langle d_1\rangle$, or equivalently
$\tau \sim 1/N$. 
The sampling of the energy
bands by discrete Bloch numbers $q$ is then too coarse to reveal
that continuous bands are underlying the discrete levels,
and the full spectrum appears as a superposition of $N$ independent
spectra. 

For $g_1/N^2 < 1$, on the other hand,  we have $g_1 \tau < N$ throughout the
range of validity of Eq.~(\ref{eq:result}) 
($0<\tau\stackrel{<}{\sim}1/N$)
and the sum in (\ref{eq:result}) 
can be restricted to $\mu=0$. Classically this means that the diffusion
cloud generated by the propagator of winding numbers is still well localized
at times corresponding to the Heisenberg time of the unit cell
($\tau \sim 1/N$).
Semiclassically, the coarse-grained diffusion
propagator encodes the quantum-mechanical spectral correlations due to the
existence of energy bands. At times of the order of
$\tau \sim 1/N$, where individual
bands begin to be resolved, the spectrum appears to be $N$-fold degenerate.
As anticipated in the introduction,
the clustering of the levels into bands is reflected in the spectral
form factor: at $\tau \sim 1/N$, 
the form factor exhibits a peak of the order of $N$,
$S_N(0;\tau)\sim N$.
Hence, for $g_1/N^2< 1$, the semiclassical expression Eq.~(\ref{eq:result})
describes the effect of the quantum levels clustering into bands
and relates this to the space-time dependence of the classical
diffusion propagator.

\begin{figure}
\centerline{\psfig{figure=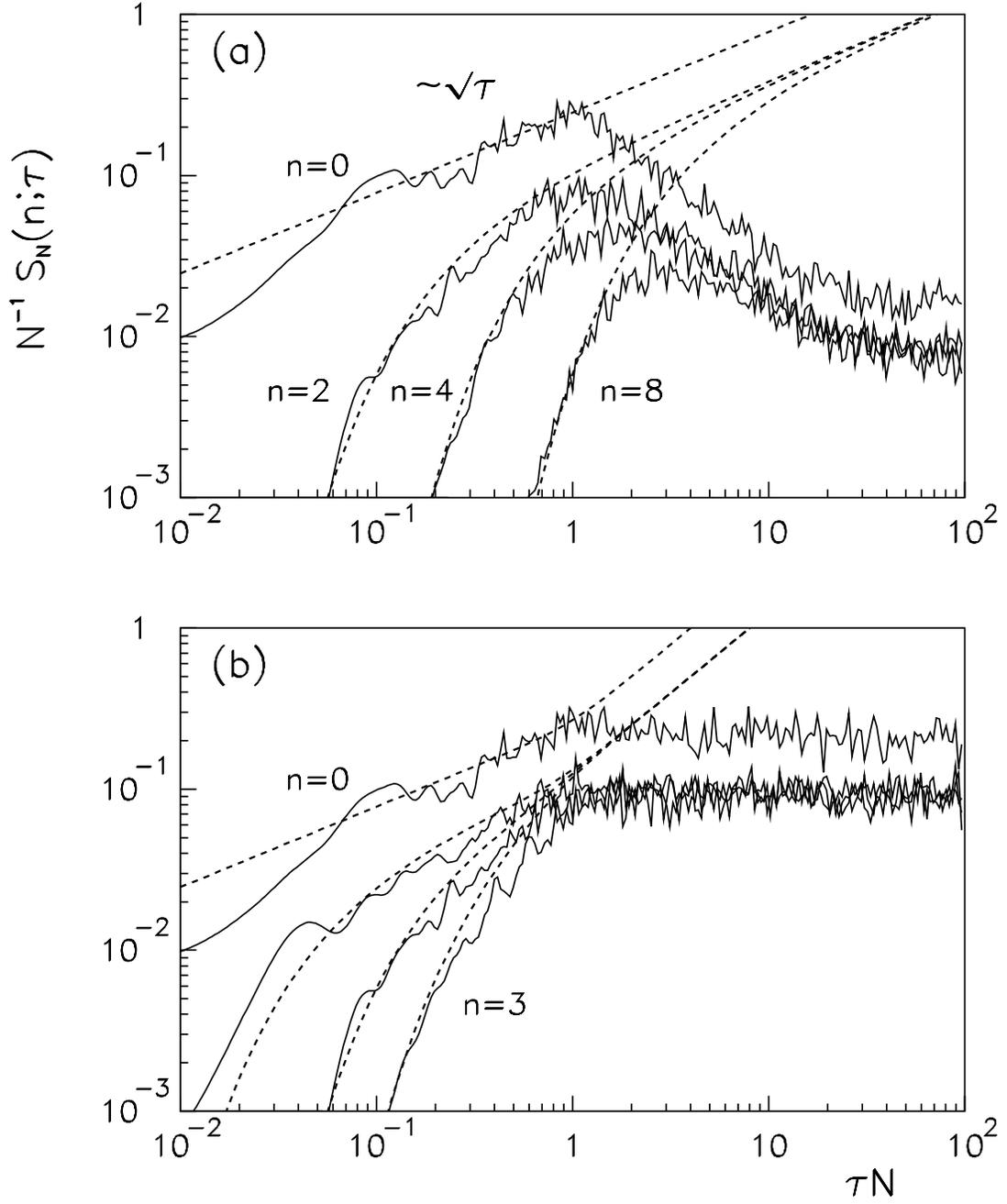,width=15cm}}
\caption{\label{fig:result} Shows $S_N(n;\tau)$ as extracted
from the quantum spectrum for a periodic chain, (a) for $N=128$
and $n=0,2,4,8$ and (b) 
for $N=8$ and $n=0,1,2,3$ (solid lines).
Also shown (dashed lines)
are the results of the semiclassical approximation Eq.~(\protect\ref{eq:result}).}
\end{figure}

In the remainder of this section, we discuss the signatures of level
clustering in numerical band spectra and compare with the predictions of
Eq.~(\ref{eq:result}). We have calculated the quantum-mechanical form
factors $S_N(n;\tau)$ for chains with $N=8$ and $N=128$ unit cells,
respectively, and with the same geometrical parameters as used in the
classical simulations described above.  Since $S_N(n;\tau)$ is not
self-averaging with respect to energy,
we have taken the
mean of seven samples with slightly 
different geometries (the relative deviations
are of the order of a few de Broglie wavelengths). In addition, we have
performed an energy average, around a mean energy of $2mE/\hbar^2
\simeq 8.9 \times 10^3$.  For unit channel width, this corresponds to
$30$ open channels.  
Fig.~\ref{fig:result}(a) shows the quantum data for
$N=128$ and winding numbers $n=0,2,4,8$ (solid lines). 
Fig.~\ref{fig:result}(b) shows the quantum data for
$N=8$ and winding numbers $n=0,1,2,3$ (solid lines). 
In both cases, the 
semiclassical approximation according to Eq.~(\ref{eq:result})
is also shown (dashed lines). We
observe reasonable agreement with the semiclassical result for times
$\tau\stackrel{<}{\sim} 1/N$. For larger times, the diagonal
approximation leading to Eq.~(\ref{eq:result}) is inappropriate as
remarked above. In the present section, we restrict our discussion to times
$\tau \stackrel{<}{\sim} 1/N$. 

For $N=128$ as well as for $N=8$, the quantum data for zero winding number
($n=0$) are
enhanced by a factor of two, in keeping with the prediction of
Eqs.~(\ref{eq:g}) and (\ref{eq:result}). 
% For $N=8$, the enhancement
% is also present for $n=4$, as expected from 
% (\ref{eq:g}) and (\ref{eq:result}).
For 
% zero winding number 
$n=0$ and for very small times $\tau \ll 1/N$, 
we observe a square-root dependence of the form factor,
$S_N(0;\tau) \sim \sqrt{\tau}$. 
This law persists up to
$\tau \sim 1/N$ 
and can be understood in terms
of the behaviour of $P(0;t)$.
For small times, $P(0;t) \sim
1/\sqrt{t}$.  In Eq.~(\ref{eq:result}) this corresponds to keeping only the
$\mu=0$ term which leads to $S_N(0;\tau)\sim \sqrt{\tau}$.  For very short 
chains, on the other hand, 
corrections due to the finite length modify this behaviour 
when the $\mu\neq 0$ terms contribute significantly.

In both cases, the conductance is
$g_1 \simeq 33$. 
In the first case ($N=128$) we have $g_1/N^2 < 1$ and the structure of
$S_N(0;\tau)$ at $\tau \sim 1/N$ (see Fig.~\ref{fig:result}(a)) 
is the signature of levels clustering into
bands, as discussed above. 
As $g_1/N^2$ grows larger, on the other hand, the effect
of clustering of levels disappears (see Fig.~\ref{fig:result}(b), where
$N=8$ and hence $g_1/N^2 \simeq 1$). The sampling of the bands 
by discrete Bloch numbers is then too coarse to reveal the underlying
continuous bands. 

In summary we emphasize that for very small times the
quantum dynamics of extended periodic quantum systems is governed
by diffusion. For small $g_1/N^2$, the spectral form factor exhibits
a peak $\sim N$ associated with the clustering of levels into bands.

\begin{figure}
\centerline{\psfig{figure=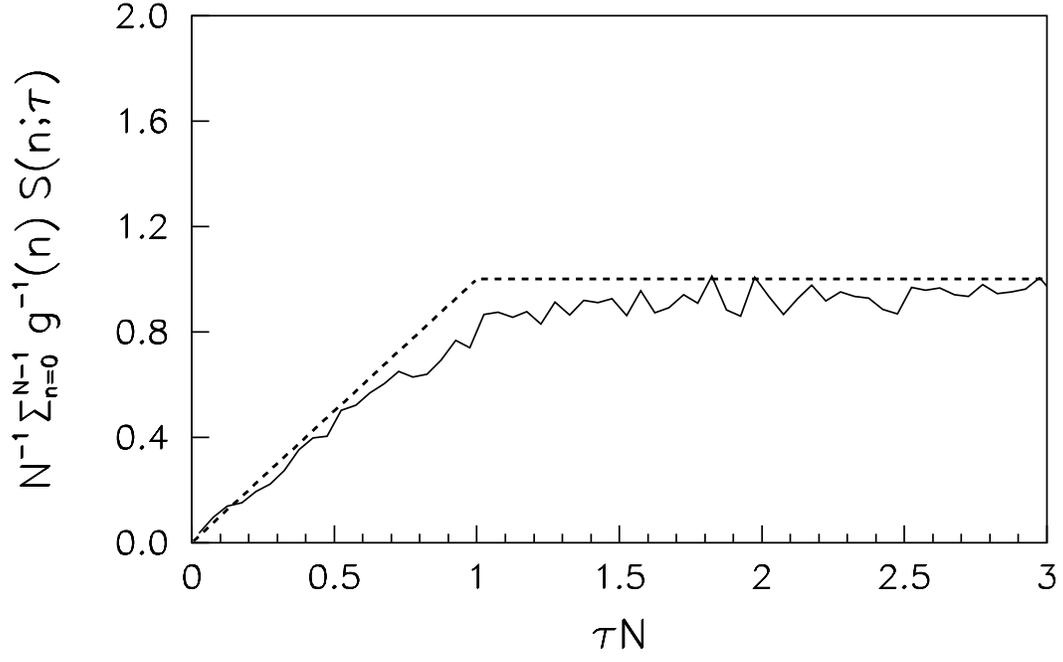,width=15cm}}
\caption{\label{fig:sum} Semiclassical sum rule.
The solid line is the sum of the spectral form factors
$S_N(n;\tau)$, while the dashed line is the semiclassical
prediction according to Eq.~(\protect\ref{eq:sumrule}).}
\end{figure}

To conclude we mention a sum rule for the spectral form factors.
According to appendix~C, the diffusion propagator 
$P(n;t)$ is normalized to
$\sum_{n=0}^{N-1} P(n;t) \simeq 1$. Using 
\begin{displaymath}
\frac{1}{N} \sum_{n=0}^{N-1} {\gamma_n}^{-1} S_N(n;\tau) \simeq \tau N
\sum_{n=0}^{N-1} P(n;2\pi\hbar \langle d_1 \rangle N\tau),
\end{displaymath}
this implies a sum rule for the spectral correlators $S_N(n;\tau)$ for $\tau <
1/N$. Together with (\ref{eq:asympt}) we have for large $N$
\begin{equation}
\frac{1}{N} \sum_{n=0}^{N-1} {\gamma_n}^{-1} S_N(n;\tau) \simeq 
\left\{
\begin{array}{ll}
\tau N & \mbox{for $\tau \stackrel{<}{\sim} 1/N$}\\
1      & \mbox{for $\tau \gg 1\,.$}
\end{array}
\right .
\label{eq:sumrule}
\end{equation}
\noindent 
This sum rule is shown in Fig.~\ref{fig:sum}. Also shown is the expectation
according to  Eq.~(\ref{eq:sumrule}). 

%%%%%%%%%%%%%%%%%%%%%%%%%%%%%%%%%%%%%%%%%%%%%%%%%%%%%%%%%
% 
% Scaling
% 
%%%%%%%%%%%%%%%%%%%%%%%%%%%%%%%%%%%%%%%%%%%%%%%%%%%%%%%%%
\section{Ballistic spreading and the scaling of the form factor}
\label{sec:scaling}
\noindent
In sections~\ref{sec:corr}
and \ref{sec:scl}, the behaviour of the spectral form
factor $S_N(n;\tau)$ has been determined for very large times
and for small times. In the first case, for times
of the order of the inverse intra-band spacing,
we have $S_N(0;\tau) \sim \gamma_n$, corresponding
to quasi-periodic motion. In the second case,
for $\tau\stackrel{<}{\sim} 1/N$, the quantum dynamics
is diffusive. In this regime, and for $g_1/N^2 < 1$, spectral 
correlations due to the clustering of levels 
lead to a pronounced peak in $S_N(0;\tau)$,
at times of the order of $\tau \sim 1/N$.
How are these two regimes to be connected?
In the present section we develop a theory for the
spectral form factor at intermediate times.

The derivation of Eq.~(\ref{eq:result}) is based on the diagonal approximation
which is known to fail for times $\tau > 1/N$. 
In the limit of large $N$, on the other hand,
the sum over $q$ in Eq.~(\ref{eq:S3}) can be replaced
by an integral
\be
\label{s-int}
S_N(n; \tau) \simeq  \frac{1}{\Delta x} 
\left|\frac{L}{2\pi}\int\!{\rm d}q\sum_\alpha e^{i nqa-2\pi i x_\alpha(q)\tau}
f(x-x_\alpha(q))\right|^2\,.  \ee According to (\ref{eq:xdef}) we have $2\pi
x_\alpha(q)\tau\sim N\tau$ and hence the phase of the integrand is rapidly
oscillating for large $\tau$.  In this case, the
integral can be evaluated in the  
saddle-point approximation.
The condition of stationary phase reads $2\pi x^\prime_\alpha(q_j) = na/\tau$
or, in terms of the unscaled energies $E_\alpha(q)$,
\begin{equation}
2\pi \langle d_1\rangle N\;\frac{\partial E_\alpha}{\partial q}
\mbox{\rule[-6mm]{0.25pt}{11mm}\raisebox{-5mm}{\hspace{3pt}$q_j$}}
={na\over \tau}\,.
\label{eq:statcond}
\end{equation}
\noindent We emphasize that the saddle points $q_{j}$ depend
on the parameters $\tau, n$ and $N$ only through the combination
${n/N\tau}$.
The resulting expression for $S_N(n; \tau)$ is
$$
S_N(n; \tau) \simeq {1\over \Delta x}
\left|\frac{L}{2\pi}\sum_\alpha \sum_{q_j}
{f(x-x_\alpha(q_j)) \over \sqrt{x''_\alpha(q_j)\tau}} 
e^{i nq_j a-2\pi i x_\alpha(q_j)\tau}\right|^2\,.
$$
The sum over $\alpha$ and $q_{j}$ involves many terms with uncorrelated
phases. Assuming some smoothing over $\tau$, we can justify replacing it by
\begin{equation}
\label{scaling}
S_N(n; \tau)={\gamma_n\over \tau}
F\left({n\over N\tau}\right)\,,
\end{equation}
where
$$
F\left({n\over N\tau}\right) 
\simeq {a^2\over 4\pi^2}   \frac{N}{\Delta x}
\sum_\alpha \sum_{q_j}
{f^2(x-x_\alpha(q_j))\over \langle d_1\rangle E_\alpha''(q_j)}.
$$
\noindent In writing (\ref{scaling}),
we have used the fact that the saddle points are functions of ${n/N\tau}$.
For $n=0$, in particular, Eq.~(\ref{scaling}) implies $S_N(0;\tau) \sim 1/\tau$.
According to Eq.~(\ref{eq:statcond}), this behaviour
corresponds to a ballistic quantum dynamics,
$\sqrt{\langle n^2\rangle}\sim \tau$.  This is in
contrast to the diffusive spreading derived for
short times, $\langle n^2\rangle\sim\tau$ for small $\tau$.

Clearly, the ballistic spreading is limited by the finite
system size. We denote the time it takes for
the distribution in $n$ to spread over the entire
system by $\tau^\ast$. 
This is a time of the order 
of the inverse intra-band spacing.
According to the discussion in section~\ref{sec:corr}, $\tau^\ast$ is just
the time where the form factor $S_N(n;\tau)$ saturates at its
asymptotic value. This saturation, discussed above in section~\ref{sec:cross},
is not described by Eq.~(\ref{scaling}).

\begin{figure}
\centerline{\psfig{figure=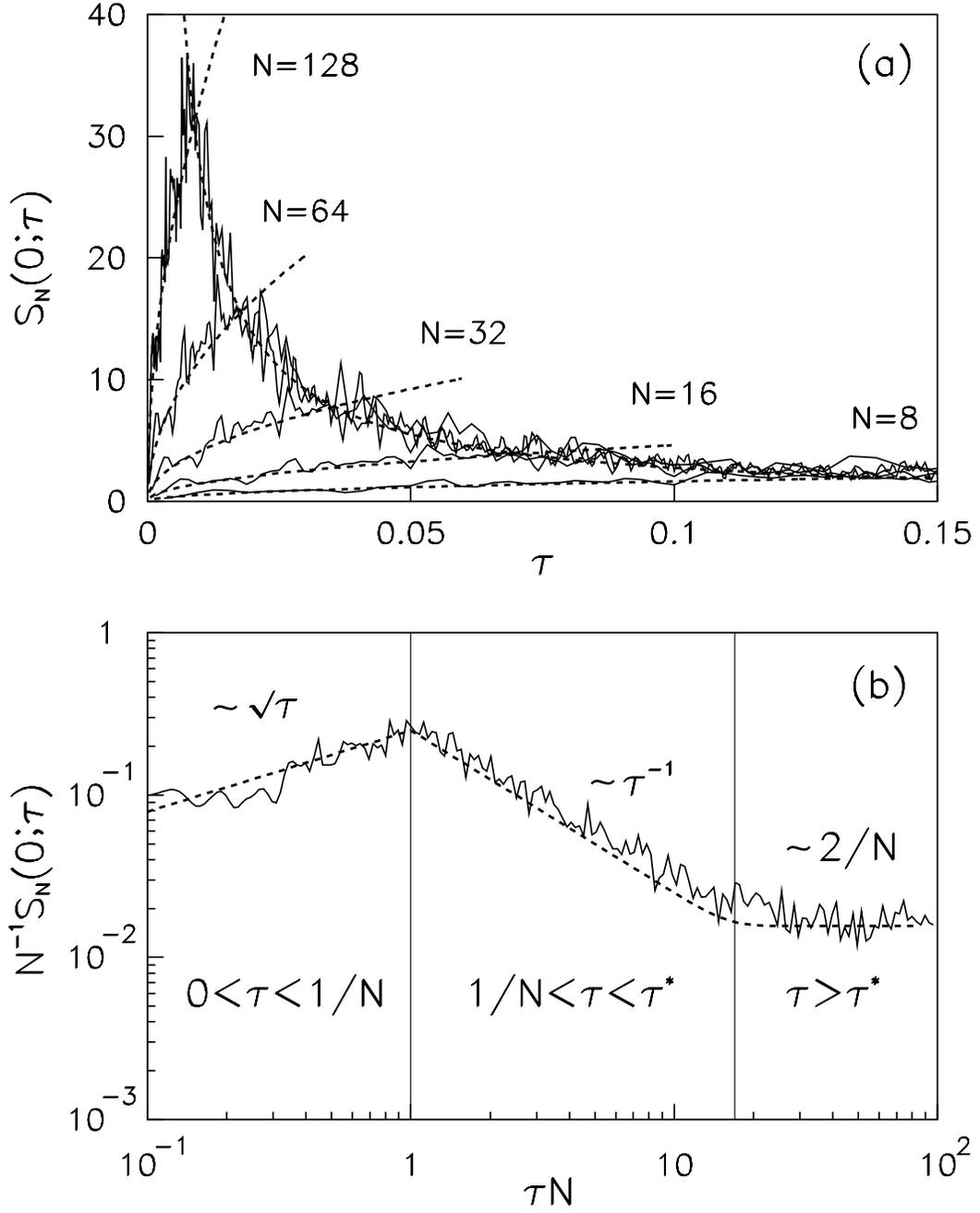,width=15cm}}
\caption{\label{fig:scaling} Scaling of $S_N(n;\tau)$.
(a) shows the scaling of $S_N(n;\tau)$ with $N$,
for $N=8,16,32,64$ and $N=128$. (b) shows the various
regimes in a doubly logarithmic plot: {(i)} the
diffusive increase $\protect\sim\protect\sqrt{\tau}$ for small times, 
{(ii)}
the peak $\protect\sim N$ at $\tau = 1/N$ due to the level clustering,
{(iii)} the $1/\tau$ decay associated with ballistic quantum
spreading and finally {(iv)} the constant asymptote
at the time scale corresponding to the
mean intra-band spacing due to quasi-periodic motion.}
\end{figure}

For $n=0$, Eq.~(\ref{scaling})
takes a particularly simple form. Matching
(\ref{scaling}) at $\tau=1/N$ with the peak derived in the diagonal
approximation, we find
\be\label{decay}
%F(0) = {a\over 2\pi}{1\over \sqrt{\hbar \langle d_1\rangle D}}\,.
F(0) = 1/\sqrt{2 g_1}\,.
\ee
\noindent Furthermore, since $S_N(0;\tau^\ast) \simeq 2$, we have
$\tau^\ast \simeq F(0)/2$.
The behaviour of $S_N(0; \tau)$ for $g_1/N^2 < 1$ is now completely  determined on
all time scales. We have
\begin{equation}
S_N(0;\tau) \sim 2
\left\{
\begin{array}{ll}
F(0)\,N\sqrt{\tau N} & \mbox{ for $0 < \tau < 1/N$}\\
F(0)\tau^{-1}        & \mbox{ for $1/N \stackrel{<}{\sim} \tau < 
F(0)/2$}\\
1               & \mbox{ for $\tau > F(0)/2\,.$}
\end{array}
\right .\label{eq:S0}
\end{equation}
This prediction is tested numerically in
Fig.~\ref{fig:scaling}. $S_N(0;\tau)$ is 
shown for
various values of $N$ with otherwise unchanged parameters in (a) and
for one particular value of $N$ on a doubly logarithmic scale in (b). The
dashed lines show the prediction of (\ref{eq:S0}).
Fig.~\ref{fig:scaling}(a) shows that the $N$-dependence of the quantum
data are adequately described by Eq.~(\ref{eq:S0}).
In Fig.~\ref{fig:scaling}(b), one can distinguish several time regimes,
as expected from Eq.~(\ref{eq:S0}):
First, for very short times
the form factor increases according to $\sim\sqrt{\tau}$,
as predicted by Eq.~(\ref{eq:S0}), 
which reflects the underlying classical diffusion.  
This behaviour continues up to $\tau\sim 1/N$. 
Second, the peak in $S_N(0;\tau)$ at $\tau\sim 1/N$ is
a signature of the levels clustering into bands.
Accordingly, the height of the peak scales as $\sim N$.
Third, for larger times, $S_N(0;\tau)$ decays as $\sim1/\tau$. 
This is consistent with a ballistically spreading quantum
distribution,
$\sqrt{\langle n^2\rangle}\sim\tau$. 
In this regime the curves corresponding to different values of $N$
coincide (see Fig.~\ref{fig:scaling}(a)) as predicted by Eq.~(\ref{eq:S0}).
Fourth, the form factor saturates at $\tau^\ast$.

As $N \to \infty$, the spectral form factor approaches a universal function. The
diffusive domain $0< \tau < 1/N$  shrinks and for $\tau \le \tau^\ast$ the
form factor is dominated by the $1/\tau$ behaviour which reflects the
formation of continuous bands. This behaviour was not discussed
before, and is one of the
most important results of the present work. 

We should finally note that periodic
systems in $d$ dimensions will exhibit a $\tau^{-d}$ behaviour, as can be easily
seen from the derivation above.

%%%%%%%%%%%%%%%%%%%%%%%%%%%%%%%%%%%%%%%%%%%%%%%%%%%%
%
% Conclusions
%
%%%%%%%%%%%%%%%%%%%%%%%%%%%%%%%%%%%%%%%%%%%%%%%%%%%%
\section{Conclusions}
\label{sec:conclusions}
In the present article, we have investigated the implications of a discrete translation
symmetry for the spectral correlations of 
quasi one-dimensional quantum systems with chaotic classical dynamics.
The decisive new feature here, as compared to disordered systems, is the presence of discretized
energy bands. We have introduced a generalized spectral form factor involving
spectral correlations both for identical as well as for  non-identical 
Bloch numbers. We have shown in detail how the time-dependence of
this form factor reflects the quantum dynamics in the periodic system:

(i) For small times, the form factor was shown to be
semiclassically equivalent to the coarse-grained classical
propagator. 
In the present case, chaotic scattering implies classical diffusion in the
extended system. This enabled us to derive an 
explicit expression for the form factor, depending only on the
diffusion constant and the system size.

(ii) At $\tau \sim 1/N$, 
the spectral form factor exhibits a
peak due to level clustering provided the discretization by the Bloch
numbers is sufficiently fine in order to reveal the underlying smooth bands. 
The height of the peak corresponds to the number of strongly correlated
levels and hence it is proportional to $N$.

(iii) For $1/N \stackrel{<}{\sim} \tau <\tau^\ast$, 
we have derived a universal scaling law for the form factor.
This regime corresponds to ballistic quantum dynamics and is
restricted by the time $\tau^\ast$ needed to ballistically cover the entire system.

(iv) For $\tau > \tau^\ast$, the form factor attains
its asymptotic constant value. This regime corresponds to
quasi-periodic motion.

We emphasize once again that our results
---  tested numerically using the example of a quasi
one-dimensional periodic chain of chaotic billiards --- are more generally
valid (see also \cite{DMSS96a}) and in fact universal. 
Finally we note that
the transition from a periodic chain to a quasi one-dimensional
disordered system is also governed by a universal law. A study concerned with
the effect of introducing disorder in the present framework
is under way.

%%%%%%%%%%%%%%%%%%%%%%%%%%%%%%%%%%%%%%%%%%%%%%%%%%%%%%%%%%%%%%%%%%%%%%
\section*{Acknowledgements} 
\noindent 
The research reported in this work was supported by grants from the US
Israel Binational Science Foundation and the Minerva Center for
Nonlinear Physics. Three of us (TD, BM and HS) would like to thank the
the Weizmann Institute of Science, Rehovot (Israel), for the kind 
hospitality during several visits. BM would like to thank the Max Planck Institute
for Physics of Complex Systems, Dresden (Germany), for financial support.

\begin{appendix}
%%%%%%%%%%%%%%%%%%%%%%%%%%%%%%%%%%%%%%%%%%%%%%%%%%%%%%%%%%%%%%%%%%%%%%
\section{The $S$-Matrix of the Unit Cell}
\label{sec:smat}
\begin{figure}
\centerline{\psfig{figure=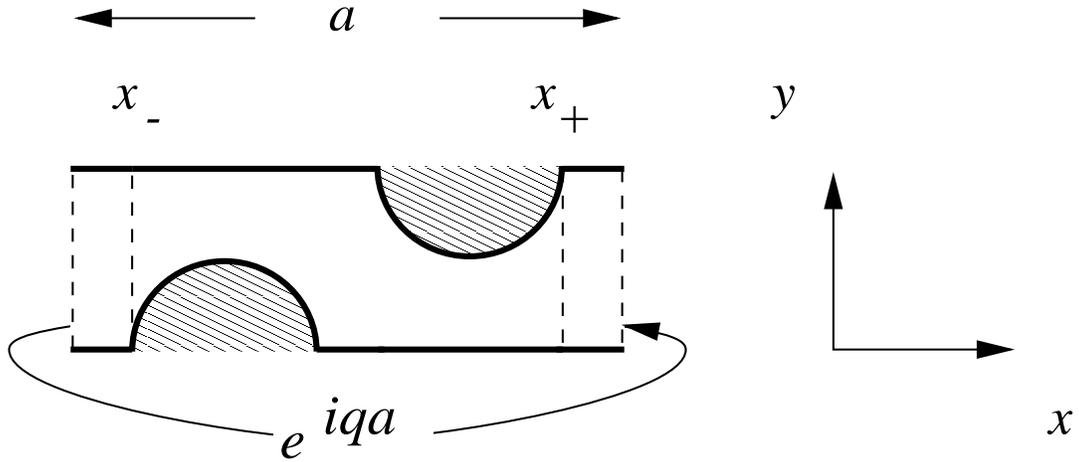,angle=270,width=15cm}}
\caption{\label{fig:unitcell} Unit cell of the periodic chain
of chaotic billiards. The unit cell has unit width and a length $a$. 
According to Bloch's theorem, the
periodic chain can be quantized using
Eq.~(\protect\ref{eq:seceq}). This corresponds to quantizing a single
unit cell with periodic boundary conditions and an additional phase $\exp(i q
a)$.}
\end{figure}

\noindent
In this appendix we briefly describe the numerical computation of the
generalized $S$-matrix of the unit cell
needed according to (\ref{eq:seceq}) for the
quantization of the periodic billiard chain. 
As shown in Fig.~\ref{fig:unitcell}, the unit cell consists
of two half discs with radii $R_1$ and $R_2$.
For the method we
employ here, it is essential that the projections of the half discs onto the
$x$-axis do not overlap (see Fig.~\ref{fig:unitcell}). This allows us
to express the
total $S$-matrix in terms of the reflection and transmission matrices of
a single half disc in a waveguide using the concatenation formulae
\bea
\begin{array}{l}
t=t_2(1+r_1'[1-r_2r_1']^{-1}r_2)t_1\\
r=r_1+t_1'[1-r_2r_1']^{-1}r_2t_1\\
t'=t_1'[1-r_2r_1']^{-1}t_2'\\
r'=r_2'+t_2r_1'[1-r_2r_1']^{-1}t_2'\,.
\end{array}
\eea
Here $r_1,r_1'$ and $t_1,t_1'$ describe the scattering off the left
half disc,
while $r_2,r_2'$ and
$t_2,t_2'$ describe the scattering off the right half disc. The
definitions of these matrices are analogous to the definition of $r,r'$
and $t,t'$ in (\ref{strafo}) and (\ref{srt}). 

The computation of the $S$-matrix of a single half disc inside the
waveguide remains to be explained. We assume that its center is
located at the origin. Other reference points as well as the
reflection from $y=1/2$ (which is needed to obtain the right obstacle
in Fig.~\ref{fig:unitcell}) can be obtained in terms of unitary
transformations. Moreover we use the reflection symmetry of the half disc 
in order to decompose the $S$-matrix into a symmetric part $S^+$
and an antisymmetric part $S^-$, which are related to the reflection
and transmission matrices according to $S^\pm=t\pm r$. The calculation
of $S^-$ including evanescent modes has been described in detail in
\cite{SS95} and $S^+$ can be found by straightforward analogy. The
result can be summarized in the matrix equation
\be
S^{\pm}=1+C^\pm\,T\,(P^\pm)^{-1}\,(C^\pm)^{\rm tr}\,.
\ee
Here $T$ denotes the transition matrix for the scattering from a
circle in the free plane which is diagonal in angular momentum
representation and has the elements $T_l(kR)=-J_l(kR)/H_l^+(kR)$.
The matrix 
\be
P^\pm_{ll'}=\delta_{ll'}-[F_{l-l'}(2k)\mp F_{l+l'}(2k)]T_{l'}(kR)
\ee
with 
\begin{equation}\label{strfun}
F_l(x) = 2\sum_{n = 1}^{\infty} \cos\left(\frac{l\pi}{2}\right)H_l^+(nx) 
\end{equation}
describes the coupling of angular momenta due to the presence of the
waveguide. The angular momentum runs over all positive even numbers
for $S^-$ and over all positive odd numbers for $S^+$. For the
numerical computation of the only conditionally convergent series
(\ref{strfun}), a method has been developed in
\cite{BS89}. Finally, the matrix 
\be
C^\pm_{j,l}=i\sqrt{2\/k_j}(R_{j,l}\pm R_{j,-l})
\ee
with
\be
R_{j,l}= \left(\frac{-i\,\mbox{sgn}(l)\,k_j + j\pi}{k}\right)^{|l|}
\ee
contains the expansion coefficients of the normal mode $n$ in the 
free-channel region in the angular-momentum basis which is appropriate
close to the semicircular obstacles.
The method described in this appendix
has been successfully applied to compute $S^-$ up
to values of $k\approx 1000$ \cite{PSSU96b}, which is far beyond
what is needed within the present application.
\newpage
%%%%%%%%%%%%%%%%%%%%%%%%%%%%%%%%%%%%%%%%%%%%%%%%%%%%%%%%%%%%%%%%%%%%%%
\section{Classical mechanics of periodic chains}
\label{sec:diff}
\begin{figure}
\centerline{\psfig{figure=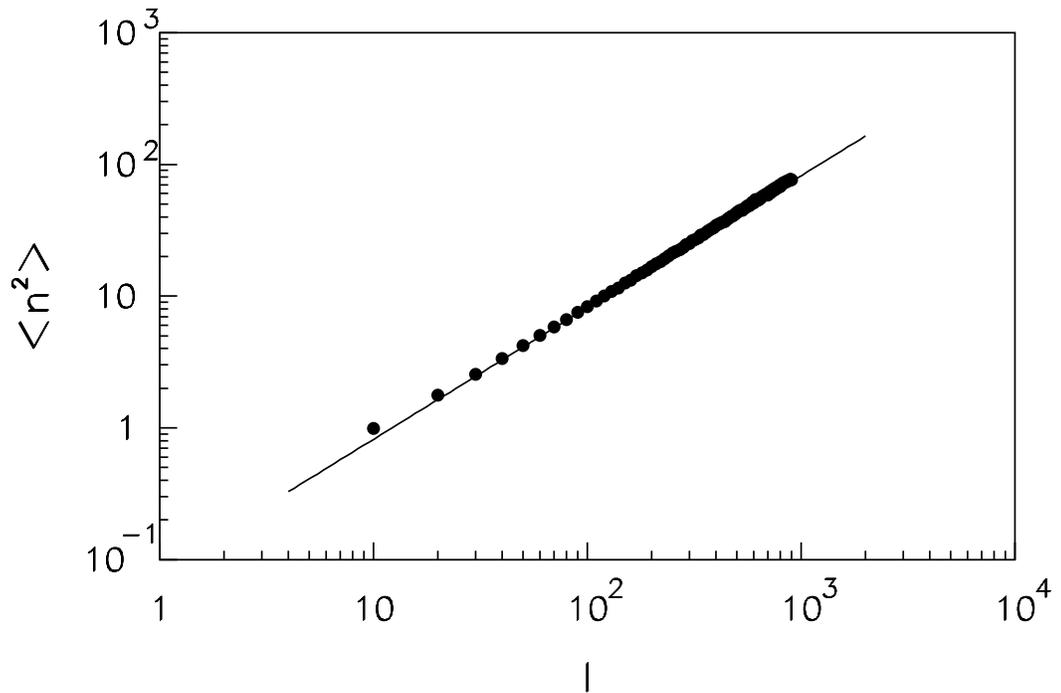,width=15cm}}
\caption{\label{fig:diff} Diffusion in the periodic chain.
Shown is the the result of a classical simulation: The variance
$\protect\langle n^2\protect\rangle$ of winding numbers $n$ for 
trajectory segments of length $l$
increases according to $\protect\langle n^2\protect\rangle \sim l$. }
\end{figure}

\begin{figure}
\centerline{\psfig{figure=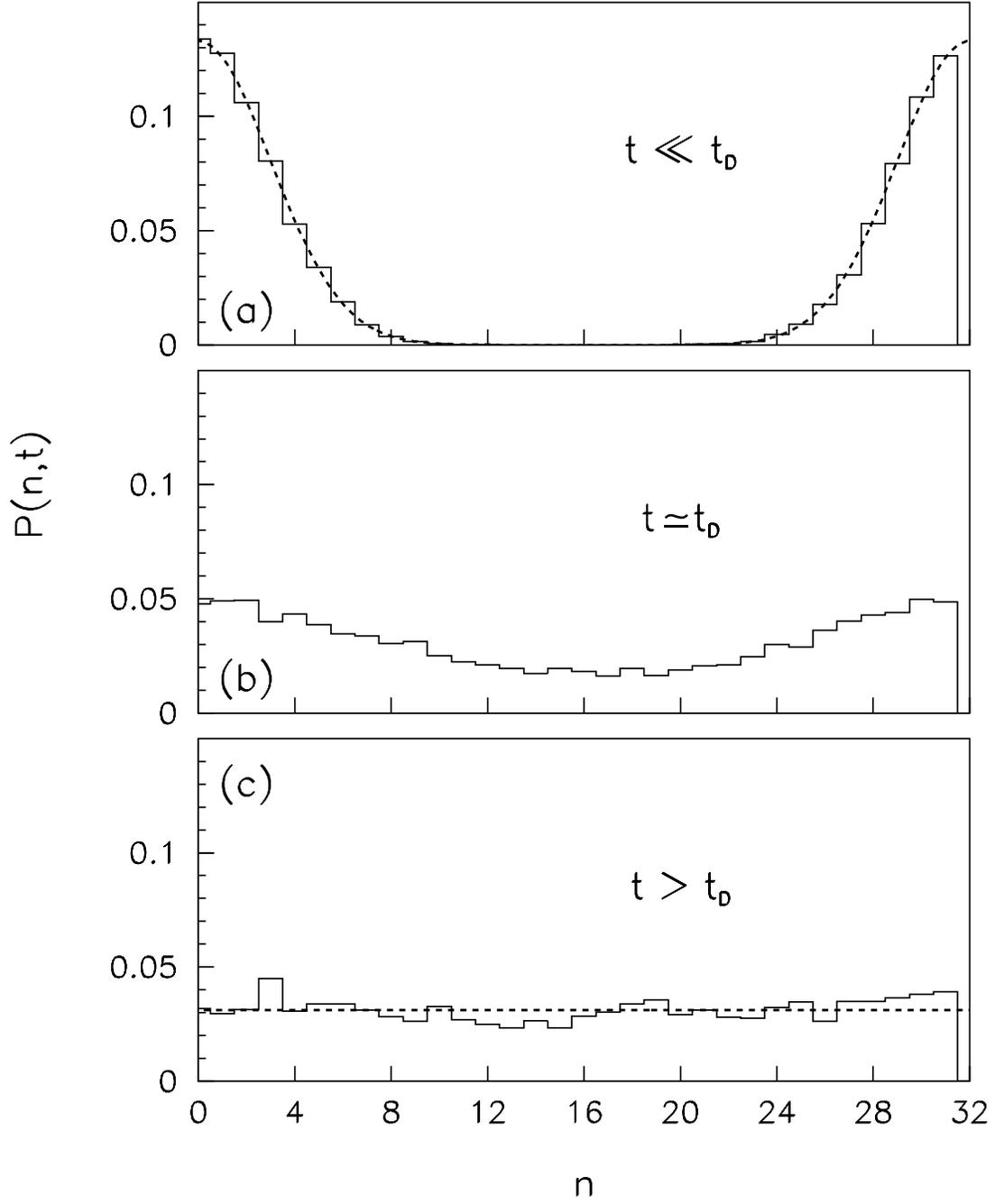,width=15cm}}
\caption{\label{fig:gaussian} 
The distribution of $P(n;t)$ for three cases: (a) for
$t \ll t_{\rm D}$, (b) for $t \simeq t_{\rm D}$ and (c) for $t > t_{\rm D}$.
The dashed lines show the limiting distributions (\protect\ref{eq:gaussian})
and (\protect\ref{eq:saturation}), for small and large times, respectively.}
\end{figure}

\noindent
In this appendix, we discuss the classical mechanics of a
periodic chain of chaotic scattering elements  with periodic boundary
conditions. We assume that the chain 
extends along the $x$-direction and consists of $N$ identical
copies of a unit cell (of length $a$).
Let ${\bf z}=(x,y,\varphi)$ denote a phase space vector 
on the energy shell ($y$ and $\varphi$ are the
transversal coordinate and the direction of the momentum,
respectively). 
The dynamics on the energy shell is governed by the
propagator
\be\label{pfo}
P({\bf z}, {\bf z'}; t)=\delta({\bf Z}({\bf z'}, t)-{\bf z})\,,
\ee
where ${\bf Z}({\bf z'}, t)$ denotes the trajectory started at $t=0$
from ${\bf z'}$. Since the unit cell is chaotic (neglecting the
possible influence of an infinite horizon), both $y$ and $\varphi$ are
effectively decorrelated on time scales larger than the 
ergodic time of the unit cell. Consequently, we obtain 
diffusive spreading in the $x$-direction
$$
\label{eq:x2Dt}
\langle x^2 \rangle \simeq D \,t\,.
$$
\noindent The classical propagator solves the diffusion equation
\be\label{diffeq}
\Bigg[\frac{\partial}{\partial t} - \frac{D}{2}
     \frac{\partial^2}{\partial x^2} \Bigg] P(x,x^\prime;t) = 0\,,
\ee
with the boundary condition $P(x;0) = \delta^{(L)}(x)$, where
$\delta^{(L)}(x)$ is the $L$-periodic delta function.
The propagator is normalized
according to $\int_0^L
\!{\rm d}x\;P(x;t) = 1$. The solution of (\ref{diffeq}) can be easily found
by representing $P(x,t)$ in terms of eigenfunctions of the stationary
problem. We obtain
\be
\label{eq:Px}
P(x;t) = \frac{1}{L} \sum_{\nu = -\infty}^\infty
\exp\left[2\pi i \nu {x\/L} - D\,\left(2\pi\nu\/L\right)^2\,{t\/2} \right]\,.
\ee
It is appropriate to introduce an additional coarse graining of the
coordinate $x$ by defining 
\be\label{cg}
n=[x/a]
\ee 
where $[\ldots]$ denotes the integer part. 
As will be seen below, $n$ can be interpreted as the
winding number of a trajectory around the unit 
cell (with periodic boundary
conditions).
>From Eq.~(\ref{eq:Px}) we have
\be
\label{eq:Pn}
P(n;t) \simeq \frac{1}{N} \sum_{\nu = -\infty}^\infty
\exp\left(2\pi i \nu {n\/N} - {D\/a^2}\,\left[{2\pi\nu\/N}\right]^2\,{t\/2} \right)\,,
\end{equation}
which after Poisson summation can be written as
\begin{eqnarray}\label{psum}
P(n;t) &=& {a\/\sqrt{2\pi D\,t}} \sum_{\mu = -\infty}^\infty
\exp\left(-{a^2\/2\,D\,t}[n-\mu N]^2\right)\,.
\end{eqnarray}
Due to the finite system size, the propagator $P(n;t)$ exhibits a
crossover from diffusion to equidistribution on the time scale of the
Thouless time $t_{\rm D} = L^2/\pi D$. 
Indeed for small times $t\rightarrow 0$, only the $\mu=0$ term contributes
in (\ref{psum}) and we have
\begin{equation}
\label{eq:gaussian}
P(n;t) \simeq \frac{a}{\sqrt{2\pi\,D\,t}}
\exp\left({-\frac{n^2a^2}{2\,D\,t}}\right)\,.
\end{equation}
For large times ($t\gg t_{\rm D}$) the dominant term in (\ref{eq:Px})
is $\nu=0$ and $P(n;t)$ saturates at
\begin{equation}
\label{eq:saturation}
P(n;t)\simeq \frac{1}{N}\,.
\end{equation}
\noindent This behaviour is summarized in Figs.~\ref{fig:diff}
and \ref{fig:gaussian}. Fig.~\ref{fig:diff} shows the 
diffusion of winding numbers. The solid dots are a numerical
simulation for an infinite periodic chain. The solid line
is a fit to the diffusion law $\langle x^2\rangle = D t$, or
equivalently $\langle n^2 \rangle = D m/\hbar k a^2\; l$.
We obtain $D m/\hbar k a^2 \simeq 0.082$. Fig.~\ref{fig:gaussian} shows $P(n;t)$
as a function of $n$ for three cases: (a) for $t \ll t_{\rm D}$,
(b) for $t \simeq t_{\rm D}$ and (c) for $t > t_{\rm D}$.

So far we have derived a diffusion equation for the coarse-grained
propagator $P(n;t)$ by neglecting all details of the chaotic dynamics
within the unit cell. In the following we derive an alternative
expression for $P(n;t)$ in terms of all periodic orbits of the unit cell, 
which will later allow to make contact with the semiclassical
spectral density. 
For this purpose we employ (\ref{pfo}) and
(\ref{cg}) to write the coarse-grained propagator as 
\be
P(n;t)={1\/\Omega_1(E)}\int_{\Omega_1(E)} {\rm d}{\bf z}\,{\rm d}{\bf z'}
P({\bf z}+n{\bf a},{\bf z'};t)\,,
\ee
where ${\bf a}=(a,y=0,\varphi=0)$ and
$\Omega_1(E)$ denotes the volume of the energy shell for 
motion within the unit cell 
(as opposed to the volume of the energy shell
for the entire chain).
Since the dynamics within the unit cell is chaotic,
the transition probability is
effectively independent of $\bf z'$, and we use this to replace $\bf
z'$ by $\bf z$:
\bea\label{pptr}
P(n;t)&\simeq&{1\/\Omega_1(E)}\int_{\Omega_1(E)} \!\!\!\!{\rm d}{\bf z}\,{\rm d}{\bf z'}
P({\bf z}+n{\bf a},{\bf z};t)\nonumber\\
&=& \int_{\Omega_1(E)} \!\!\!\!{\rm d}{\bf z}\,P({\bf z}+n{\bf a},{\bf z};t)\,.
\eea
We have now expressed the coarse-grained popagator by the probability to
return to the same phase-space point up to a translation by $n$ lattice
vectors or, when interpreted in the unit cell, as the joint probability to
return and to have completed $n$ windings around the cylinder.
The total probability to return on a compact phase space can be expressed as a
sum over all periodic orbits \cite{Dit96} and an analogous relation for
$P(n;t)$ can be obtained in terms of the periodic orbits of the unit cell with
winding number $n$ (disregarding bouncing ball orbits). We insert (\ref{pfo})
into (\ref{pptr}) and use a linear approximation for ${\bf Z}({\bf z}, t)$ in
the vicinity of each periodic orbit $p$.  $x$ and $y$ are replaced by local
coordinates $x_\bot$ and $x_\|$ aligned with the orbit, where the latter is
cyclic with the length $L_p$ of the orbit $x_\|+L_p\equiv x_\|$. Finally we
have
\bea\label{eq:retp}
P(n;t)&\simeq& \sum_{p,r}\int_{\Omega_1(E)} {\rm d}x_\|\,{\rm d}x_\bot{\rm
d}\varphi\,\delta^{(N)}_{n,rn_p}\;
\delta^{(L_p)}(L_p+2k[t-rT_p])\,\nonumber\\&\times&
\delta\left({\partial x_\bot(t)\/\partial x_\bot(0)}\Delta x_\bot+
{\partial x_\bot(t)\/\partial\varphi(0)}\Delta\varphi-\Delta x_\bot\right)\,
\delta\left({\partial\varphi(t)\/\partial x_\bot(0)}\Delta x_\bot+
{\partial\varphi(t)\/\partial\varphi(0)}\Delta\varphi-\Delta\varphi\right)\nonumber\\
&=& \sum_{p,r}\,\delta^{(N)}_{n,rn_p}\,
{T_p\/|\det(M_p^r-1)|} \,\delta(t-rT_p)\,,
\eea
where $T_p$ denotes the period of the periodic orbit and $M_p$ is the
monodromy matrix.  The sum over $r$ is a sum over repetitions of primitive
periodic orbits $p$. In this way we have expressed the coarse-grained
classical propagator of the extended billiard $P(n,t)$ using periodic orbits
of the unit cell. We make use of this result in the semiclassical analysis of
the billiard in section \ref{sec:scl}.
\end{appendix}

%%%%%%%%%%%%%%%%%%%%%%%%%%%%%%%%%%%%%%%%%%%%%%%%%%%%%%%%%%%%%%%%%%%
%
% Bibliography
%
%%%%%%%%%%%%%%%%%%%%%%%%%%%%%%%%%%%%%%%%%%%%%%%%%%%%%%%%%%%%%%%%%%%%
%\bibliographystyle{unsrt}
%\bibliography{qchaos}

%%%%%%%%%%%%%%%%%%%%%%%%%%%%%%%%%%%%%%%%%%%%%%%%%%%%%%%%%%%%%%%%%%%
%
% Figures
%
%%%%%%%%%%%%%%%%%%%%%%%%%%%%%%%%%%%%%%%%%%%%%%%%%%%%%%%%%%%%%%%%%%%%

%\newpage\psfig{figure=fig1.ps}
%\newpage\psfig{figure=fig2.ps,angle=270}
%\newpage\psfig{figure=fig3.ps,angle=270}
%\newpage\psfig{figure=fig4.ps}
%\newpage\psfig{figure=fig5.ps}
%\newpage\psfig{figure=fig6.ps}
%\newpage\psfig{figure=fig7.ps}
%\newpage\psfig{figure=fig8.ps}
%\newpage\psfig{figure=fig9.ps}
%\newpage\psfig{figure=fig10.ps}
%\newpage\psfig{figure=fig11.ps}
%\newpage\psfig{figure=fig12.ps}
%\newpage\psfig{figure=fig13.ps}
\end{document}